\documentclass[11pt]{article}
\usepackage{latexsym} 
\usepackage{graphicx}
\usepackage{amsfonts}
\usepackage{amsmath}
\usepackage{amsthm}
\usepackage{amssymb}

\renewcommand{\theequation}{\thesection.\arabic{equation}}
\newcounter{subequation}[equation]
\makeatletter

\newcommand{\p}{^{\prime}}
\newcommand{\pp}{^{\prime \prime}}
\newcommand{\alphadot}{\stackrel{\cdot}\alpha}
\newcommand{\betadot}{\stackrel{\cdot}\beta}

\expandafter\let\expandafter\reset@font\csname reset@font\endcsname

\def\subeqnarray{\arraycolsep1pt
    \def\@eqnnum\stepcounter##1{\stepcounter{subequation}%
        {\reset@font\rm(\theequation\alph{subequation})}}
\jot5mm     \eqnarray}

\makeatother

\def\be{\begin{equation}}

\def\ee{\end{equation}}
\def\bea{\begin{eqnarray}}
\def\eea{\end{eqnarray}}
\def\ba{\begin{array}}
\def\ea{\end{array}}
\def\dd{\partial}

\def\e {\varepsilon}
\def\half{\frac{1}{2}}

\def\one#1{#1^{\raise5pt\hbox{$\scriptstyle\!\!\!\!1$}}\,{}}
\def\two#1{#1^{\raise5pt\hbox{$\scriptstyle\!\!\!\!2$}}\,{}}

\def\tilde{\widetilde}
\def\II{\hbox{{1}\kern-.25em\hbox{l}}}

\makeatletter
\def\binrel@#1{\begingroup
  \setboxz@h{\thinmuskip0mu
    \medmuskip\m@ne mu\thickmuskip\@ne mu
    \setbox\tw@\hbox{$#1\m@th$}\kern-\wd\tw@
    ${}#1{}\m@th$}%
  \edef\@tempa{\endgroup\let\noexpand\binrel@@
    \ifdim\wdz@<\z@ \mathbin
    \else\ifdim\wdz@>\z@ \mathrel
    \else \relax\fi\fi}%
  \@tempa
}
\let\binrel@@\relax
\def\overset#1#2{\binrel@{#2}%
  \binrel@@{\mathop{\kern\z@#2}\limits^{#1}}}
\def\underset#1#2{\binrel@{#2}%
  \binrel@@{\mathop{\kern\z@#2}\limits_{#1}}}
\makeatother
\newfont{\bbd}{msbm10 scaled\magstep1}

\def\R{\hbox{\bbd R}}
\def\Q{\hbox{\bbd Q}}

\begin{document}

\vspace*{1cm}

\begin{center}
{\LARGE \bf{  Yangian symmetry applied to Quantum chromodynamics} }

\vspace{0.5cm}

{\large \sf

R. Kirschner\footnote{\sc e-mail:Roland.Kirschner@itp.uni-leipzig.de}} 

\vspace{0.5cm}

{\it Institut f\"ur Theoretische
Physik, Universit\"at Leipzig, \\
PF 100 920, D-04009 Leipzig, Germany}

\end{center}

\vspace{.5cm}
\begin{abstract}
\noindent
We review applications of Yangain symmetry to high-energy QCD
phenomenology. Some basic facts about high-energy QCD are recalled,
in particular the spinor-helicity form of scattering amplitudes, the scale
evolution equations of deep-inelastic scattering structure functions and the
high-energy asymptotics of scattering. 
As the working tool the Yangian symmetric correlators are introduced 
and constructed in the framework of the Yangian algebra of $g\ell(n)$ type.
We present the application to the tree scattering amplitudes and their iterative
relation, to the parton splitting amplitudes and the kernels of the
scale evolution equations of structure functions and to the equations
describing the high-energy asymptotics of scattering.   

\end{abstract}

{\bf keywords:} Non-abelian gauge field theory, Quantum chromodynamics, 
scattering amplitudes, Bjorken asymptotics, Regge asymptotics, 
Yang-Baxter relations, Yangian algebra, Yangian symmetric correlators.

\vspace{2cm}

{\small \tableofcontents}
\renewcommand{\refname}{References.}
\renewcommand{\thefootnote}{\arabic{footnote}}
\setcounter{footnote}{0} \setcounter{equation}{0}

\section{Gauge field theories and Quantum integrable systems}
 \setcounter{equation}{0}

The history of Quantum integrable systems  started with simple models 
for demonstrating basic physics, spin chains \cite{Bethe},
scattering in one dimension \cite{CNY}, two-dimensional lattices
\cite{Baxter}.  In the first period the application to Quantum field theory
was restricted to exceptional and sophisticated models in 1+1 dimensions.

The basic methodical concepts, Yang-Baxter relations, Bethe ansatz,
established in the first basic publications, have been developed and extended to
the Quantum inverse scattering method \cite{FST, TTF, KuSk1, Fad, KBI93}. 
A new chapter of mathematics, the Quantum algebra,  has been opened
\cite{Drinfeld, Woronowicz}, which is actively developing. 

Quantum chromodynamics (QCD) is now well established as the gauge field
theory of hadronic interactions, relying on basic concepts as
non-abelian gauge field theory \cite{Yang54}, its quantization
\cite{FP67, LDF69}, renormalizability \cite{tHooft}, asymptotic freedom
\cite{GW73, P73}, and the quark model \cite{GellMann}. After the period of
experimental confirmation  the phenomenology of 
high-energy scattering is now oriented on finding signals beyond the 
Standard Model and on understanding the detailed structures of hadronic
interaction, where extensive perturbative QCD calculations serve as a
working tool. 

The first application of Quantum integrable systems 
to high-energy QCD was pointed out by Lipatov
\cite{LevPadua}. Treating the Regge asympototics \cite{Regge, Froissart, Gribov}
in perturbative gauge theory \cite{L76, FKL, BL78}  he showed that the multiple
exchange of reggeized gluons can be treated as a Heisenberg
quantum spin chain with the spin $\half $ representations  at its sites
replaced by infinite-dimensional representations.   

It was clear immediately, that the Bjorken asymptotics \cite{BjP69, GL72a,
GL72b,LNL74,AP77,YLD77}
and the composite
operator renormalization have similar features of integrability \cite{BDM98}. 
In both
cases the Yangian symmetry underlying integrability is based on 
conformal symmetry, which is broken by loop corrections in the case of 
QCD. The breaking is suppressed by supersymmetry and is absent in
$\mathcal{N} = 4 $ super Yang-Mills theory.    
Much work has been devoted in the last  decades to super Yang-Mills theory
under this aspect.  Quantum integrability works in the composite operator
renormalization in the planar limit to all orders \cite{BS03} and
in the computation of scattering amplitudes and Wilson loops
\cite{DHKS08,DHP09,AH}. 

Describing the relations of high-energy QCD to Yangian symmetry we do not
intend to give a review of all relevant literature. Only a few references will
be given. We try to include sufficient details of explanations in order
to make the text readable without consulting these papers and books.

We shall recollect the subjects of high-energy QCD, scattering 
amplitudes in helicity form, Bjorken asymptotics and 
Regge asymptotics, where the applications are to be addressed.
We shall formulate well-known relations, emphasizing the conformal
symmetry,
in order to prepare the comparison with the results of the Yangian symmetry.
\cite{Levrev96},  \cite{IFL10} provide a broader introduction with a similar
orientation.

We shall not attempt to give a complete introduction with general and rigorous
definitions to the basics of Quantum algebra. For an introduction to the
particular subject of Yangian algebra we refer to \cite{Molev}.
We focus on the items needed to define the Yangian symmetric correlators
(YSC), to derive and to explain the relations used for their construction.
We shall introduce the Yangian algebra notions in a way well understandable 
by physicists. The algebra generators are constructed from underlying
Heisenberg algebra generators like in Quantum mechanics the operators of
observables are constructed from the basic position and momentum operators.

Our approach uses in part ideas and methods developed for the the 
scattering amplitude computations, \cite{AH}
Unlike the majority of related publications we do not approach the amplitudes
directly and do not work in the frame of super Yang-Mills. 

The YSC are obtained as N-point functions for Yangians of type $g\ell(n)$
in terms of expressions depending on $n$ and a set of parameters. 
In the applications 
$n$ will be specified to $4$ for amplitudes or to $2$ for evolution 
kernels of  structure functions and for reggeon interaction kernels.
The representation parameters are related in particular to homogeneity weights, which 
allow to describe the scattering states by substituting particular values.
The dependence of the amplitudes or kernels on the particle type and
helicity is derived from the analytic parameter dependence of the YSC 
by approaching particular values. 
The construction methods of YSC are related to the method of on-shell graphs
in the amplititude  construction \cite{Broedel}. 
In the original form  of the latter method \cite{AH}
no  parameters appeared. In a number of papers the deformation of
amplitude expressions by   parameters has been studied
and their eventual role for regularization of loop divergencies has been
discussed \cite{FLMPS13,FLS14,BBR14,BHLY14}. The parameters characterizing
the monodromy defining the YSC play the essential role in our approach.

\section{High-energy scattering in QCD }

\subsection{Scattering amplitudes}

We shall consider the scattering amplitudes of quarks and gluons 
contributing to high-energy hadronic processes, e.g. 
inclusive jet production, $A B \to jet + X $,  
at large values of $s = (p_A + p_B)^2$. 

The advantages of using helicity states in the perturbative calculation of
scattering amplitudes in gauge theories were known for long time
\cite{Berends87} and  widely used in the detailed calculations  
for collider phenomenology \cite {Dixon96}.
Both the  momenta and the polarization states of the scattering quarks and
gluons are expressed in terms of helicity spinors.
The spinor components of massless momenta appear as products of the components
of a left and right spinor, 
\be \label{klambda}
 k_{\alpha, \alphadot} = \sigma_{\alpha, \alphadot}^{\mu} k_{\mu} 
= \lambda_{\alpha} \bar \lambda_{\alphadot} \ee
The solutions of the massless Dirac equation can be expressed in terms of
 the Weyl spinors and they represent the helicity states of the scattering
quark. 
The polarization vectors characterizing the helicity states
of the gluons can be composed 
with  the help of reference spinors 
$ \mu_{\alpha}, \bar \mu_{\alphadot} $, the factors
of the light-like gauge vector
$ q_{\alpha, \alphadot} = \mu_{\alpha} \bar \mu_{\alphadot} $.

\be \label{epslambda}
\epsilon^+_{\alpha, \alphadot} = \frac{\lambda_{\alpha} \bar
\mu_{\alphadot}}{[\bar \mu \bar \lambda]}, \ \ \ 
\epsilon^-_{\alpha, \alphadot} = \frac{\mu_{\alpha} \bar
\alpha_{\alphadot}}{< \mu  \lambda>}.
\ee
The spinor indices $\alpha, \alphadot $ take two values and
the inner product of left-handed (undotted) spinors is abbreviated
by angular brackets and the one of right-handed (dotted) by
square brackets,
\be \label{[]}  < \mu  \lambda> = \mu_1 \lambda_2 - \mu_2 \lambda_1, \ \ \
[\bar \mu \bar \lambda] = 
\bar \mu_1 \bar \lambda_2 - \bar \mu_2 \bar \lambda_1. 
\ee

At high energies the massless case can be adopted as a  first
approximation, and then
 conformal symmetry holds at tree level.
The action of infinitesimal conformal 
transformations, originally representing the actions on  space-time
coordinates, the 
 shift $P_{\mu} $, the Lorentz  $M_{\mu \nu} $,
the special-conformal $K_{mu}$ and the dilatation $D$ transformations,
appear in the spinor helicity representation as \cite{Witten04} 
$$ P_{\alpha, \alphadot} = \lambda_{\alpha} \bar \lambda_{\alphadot}, \ \ 
M_{\alpha \beta} =  \lambda_{\alpha}  \dd_{\beta}, \ \ 
\bar M_{\alphadot \betadot} =  \bar \lambda_{\alphadot}  \bar \dd_{\betadot},
$$
\be \label{PMKD} K_{\alpha, \alphadot} = \dd_{\alpha} \bar \dd_{\alphadot}, \ \ \ 
D= \half( \lambda_{\alpha} \dd^{\alpha} + 
\bar \lambda_{\alphadot} \bar \dd^{\alphadot} +2 ).
\ee
The  Lorentz vector components 
are transformed into the spinor components as in the particular case of 
the momentum above (\ref{klambda}).

The eigenvalues of the dilatation in action on 
$\epsilon^{\pm}_{\alpha, \alphadot}, \lambda_{\alpha},  \bar
\lambda_{\alphadot} $ are the canonical scaling dimensions of the corresponding
fields in the QCD action, namely $1$ or $\frac{3}{2}$ for gluons or 
quarks. (Here $D$ measures the momentum dimension as $+1$, $[D, P_{\alpha,
\alphadot}] = P_{\alpha, \alphadot} $). 
The scattering states are characterized by the helicities, the
eigenvalues of the helicity operator $\hat h$,
\be \label{hel}
 2\hat h =  \bar \lambda_{\alphadot} \bar \dd_{\alphadot} 
-  \lambda_{\alpha} \dd_{\alpha}, \ee
commuting wth all  conformal transformations (\ref{PMKD}).
 
The spinor-helicity presentation of the conformal symmetry generators does
not have the Jordan-Schwinger form. This form is obtained
by elementary canonical transformations on the left spinor
variables,
\be \label{cantrans1} 
\begin{pmatrix}
\dd_{\alpha} \\
\lambda_{\alpha} 
\end{pmatrix}
\rightarrow 
 \begin{pmatrix}
x_{\alpha} \\
- \dd_{x\ \alpha} 
\end{pmatrix}.
\ee
The two canonical pairs of the right spinors are not changed, but we prefer
to rename them as
\be \label{cantrans2}
\begin{pmatrix}
\bar \dd_{\alphadot} \\
\bar \lambda_{\alphadot} 
\end{pmatrix}
\rightarrow 
\begin{pmatrix}
\dd_{x \ 2+ \alphadot} \\
 x_{2+\alphadot} 
\end{pmatrix}.
\ee
In terms of the resulting Witten twistor canonical pairs $x_a, \dd_a$ the
conformal symmetry generators are written in the Jordan-Schwinger form as
\be \label{Mab} M_{ab} = x_a \dd_{b}. \ee
The indices take 4 values, $a,b = 1, ..., 4$.
 The helicity operator (\ref{hel}) is included as the trace of the $4 \times 4$ matrix
of generators.
\be \label{2hx}
 2 \hat h \to x_1 \dd_1 + x_2 \dd_2 + \dd_3 x_3 + \dd_4 x_4, \ \ \
2 \hat h = (x\dd) + 2. \ee

The momenta of the incoming particles $ A \ B$ lead us to prefer in the
center-of-mass frame 
the basis of light-like vectors $ n_+,n_-, n_*, n_o$,
\be \label{pApB} p_A =  {\frac{\sqrt{s}}{2}} n_+ + \mathcal{O}(\frac{M^2}{s}), \ \ \
p_B = {\frac{\sqrt{s}}{2}} n_- + \mathcal{O}(\frac{M^2}{s}). \ee
$n_*, n_o$ span the transverse plane.
The normalization is
$ n_+ n_- = 2, n_* n_o = -2 $.
A generic Lorentz vector $k$ is expanded as
\be \label{Sudak}
 k = k^- n_- + k^+ n_+ + \kappa n_* + \kappa^* n_o . \ee
If it is light-like, $k^+ k^- = \kappa \kappa^* $,  then
it can be factorised as in (\ref{klambda}),  
$ (\lambda) = (\lambda_1, \lambda_2), (\bar \lambda) = 
(\bar \lambda_1, \bar \lambda_2) $ with
\be \label{lambdak} 
 (\lambda) =  (\sqrt{k^+}, \frac{\kappa}{\sqrt {k^+}}), \ \ \ 
(\bar \lambda) =  (\sqrt{k^+}, \frac {\kappa^*}{\sqrt {k^+}}). 
\ee
The polarisation vectors of a scattering gluons of momentum $k$ are
\be \label{epsk} \epsilon^+(k) = \bar E(k) ( \frac{n_o}{\sqrt{2}} +  \frac{k^-
}{\sqrt{2}\kappa^*} n_-) , \ \  
\epsilon^-(k) = E(k) ( \frac{n_*}{\sqrt{2}} +  \frac{k^- }{\sqrt{2}
\kappa} n_- ) \ee
with $ n_-$ chosen as the gauge vector and 
$ \bar Ek) E(k) = 1$.  
Their spinor form  factorizes as written above (\ref{epslambda}), e.g.
 \be \label{epsspin} 
\epsilon_{\alpha, \alphadot}^+(k) = const \mu^-_{\alpha}
\epsilon^+_{\alphadot}, \ \ 
(\mu )  = (n_+) =  (1,0), \ \ (\epsilon^+(k) ) = E(k) ( \frac{k^-}{\kappa^*}, 1 )
\ee 
 
A way to calculate the lowest order perturbative
contributions starts from the light-cone gauge form of the action.
Being aware of the peculiarities of this non-covariant form we see the advantage 
that the physical degrees of freedom of the scattering partons are described by
the transverse components $A^*, A$  of the gauge field  potential 
and projections $ f^*, f $ of the fermion fields in the 
QCD action after imposing the gauge
$n_- A = 0 $ and integrating in the path integral over the redundant 
components $ n_+^{\mu} A_{\mu} $ and $  n_+^{\mu} \gamma_{\mu} \psi $. 
Here we consider the Dirac fermion field $\psi$ 
decomposed with respect to a basis
of Majorana spinors $u_{a,b}, a=\pm, b=o,*$ obeying
$\gamma^{\pm} u_{\pm, b} = 0 , \gamma^* u_{a,o} = \gamma^o u_{a, *} = 0 $,
with the vector of the Dirac gamma matrices decomposed in analogy to the
momentum vector (\ref{Sudak}). 
$f^*, f $ represent the two helicities of a quark of
particular chirality and flavor. 
The field degrees of freedom of the quarks and gluons are directly 
represented by the fields, $A^*, A $, the gluons of helicities
$+1, -1$,  and  $f^*, f$, the quarks of a particular flavor and particular
chirality of helicities $+ \half, - \half$.

We write the kinetic part of the resulting light cone action for later
reference,
\be \label{S2} S_{2} = \int d^4 y \mathcal{L}_{2}(y) , \ \ \ 
\mathcal{L}_{2}(y) = -2 A^{B *}(y) (\dd_+ \dd_- - \dd \dd^*) A^B(y) +
i f^{ b*} \dd_+^{-1} (\dd_+ \dd_- - \dd \dd^*)  f^b(y). \ee
$b$ runs over the range of the fundamental gauge group representation
$b=1, ..., N_c$ and $B$ over the range of the adjoint representation, $B= 1,
..., N_c^2 -1 $ and   
$ \dd_{\pm} = \half (\dd_0 \pm \dd_3), \ \ \dd^* = \half (\dd_1 + i \dd_2 )$.

The dependence of the amplitudes on the the gauge group (color ) indices
can be separated \cite{Dixon96}. Here we focus on the 
the amplitude factors carrying the dependence on momenta and
helicities.

We label the parton scattering states by their helicities and assume that
the position $I$ in the array of arguments stands for the corresponding
momentum $k_{I}$ and helicity spinors $\lambda_I, \bar \lambda_I $.
We assume all momenta ingoing, unifying in this way amplitudes related by
crossing. In the planar partial amplitudes the dependence on the 
scattering states is cyclic. 

In the case of physical (on-shell) 
momentum values  $\bar \lambda_{\alphadot} $
is complex conjugated to $\lambda_{\alpha} $. In course of a calculation
we  allow for complex momentum values. In particular modifications of 
values depending on a complex parameter $c$, preserving the momentum conservation
condition, can be considered. The analytic structure of the
$c$ dependence of the amplitudes $M(c)$ 
is determined by the physical unitarity. Tree amplitudes result in
meromorphic functions of $c$ and as a consequence one obtains the BCFW
iteration relations \cite{BCFW}. 

Consider the $N$ particle helicity amplitude $M_{1, .., N}$ and deform the
helicity variables as
\be \label{1cN}
 \bar \lambda_1 \to \bar \lambda_1 - c \bar \lambda_N, 
\lambda_N \to \lambda_N + c \lambda_1 \ee
and $k_1(c) = \lambda_1 (\bar \lambda_1 - c \bar \lambda_N)$, $
k_N(c) = (\lambda_N + c \lambda_1 ) \bar \lambda_N $.
The deformed amplitude $M_{1^*, 2, ..., N^*} (c)$ has single poles at $c=
c_l$
corresponding to the one-particle intermediate states, where
$$ P_l(c) = P_l - c \lambda_1 \bar \lambda_N, P_l = k_1+ ...+ k_l $$
is light-like,
$ P_l(c_l)^2 = 0$. This leads to the BCFW interative relation
$$ M_{1, .., N} = 
\sum_l \sum_h M^h_{1, ...,l,P}(c_l) \frac{1}{P_l^2} M^{-h}_{P_l, l+1,
...,N}(c_l).
 $$ 
 $M^h_{1, ...,l,P}(c_l)$ is the tree amplitude of
$l+1$ particles of momenta $k_1(z_l), k_2, ..., k_l, P_l(c_l)$ and $h$ is
the helicity of the last one. $M^{-h}_{P_l, l+1, ...,N}(z_l)$ is the 
tree
amplitude of $N-l+1 $ particles with momenta
$-P_l(c_l), k_{l+1}, ..., k_N(c_l) $ and helicity $-h$ of the first one. 
Here it is assumed that the deformed amplitude vanishes at $c\to \infty$,
which holds e.g. in the gluon case for helicity values $-1$ for particle 1
and $+1$ for particle $N$.  

The deformation (\ref{1cN}) can be written by shift operators as 
$$ M_{1, ..., N}(c) = e^{-c\bar \lambda_N \bar \dd_1 } e^{c \lambda_1 \dd_N} 
M_{1, ..., N}, \ \ \ \
\bar \dd_1^{\alphadot} = \frac{\dd}{\dd\bar \lambda_{ 1 \alphadot}}, 
\dd_N^{\alpha} = \frac{\dd}{\dd\lambda_{N \alpha}} , $$
and in terms of the amplitudes with energy-momentum delta included,
 $ \tilde M_{1, ..., l} = M_{1, ..., l} \delta(\lambda_{1 \alpha_1} \bar
\lambda_{1, \alphadot_1}+ ... + \lambda_{l \alpha_l} \bar
\lambda_{l, \alphadot_l} ) $,
the particular contribution to $\tilde M_{1,...,N}$
of the channel $1,..,l \to l+1,...,N$ 
obtains the form 
\be \label{BCFW}
 \int_{\mathcal{C}_l }  \frac{dc}{c}     
\exp({-c\bar \lambda_{N \alphadot} \bar \dd_1^{\alphadot} }) 
\exp( { c \lambda_{1\alpha} \dd_N^{\alpha} }) 
\int d^4 P \delta(P^2) \tilde M^h_{1, ..., l, P} 
\tilde M^{-h}_{P, l+1, ...,N} .
\ee 
$\mathcal{C}_l$ denotes the small circle around $c_l$.

The 3-point amplitudes can be used
as the building blocks to construct all tree amplitudes.
In the Parke-Taylor form \cite{PT86} we have e.g.
$$ M(-1,-1,+1) = \frac{<1 2>^4}{<1 2> <2 3> <3 1> },
\ \ \ M(-\half. -1, +\half ) = \frac{<1 2>^3 < 2 3>}{<1 2> <2 3> <3 1>}.   
$$
The amplitudes with all helicities $h_i$ reversed are obtained by replacing
the brackets   $<...>$ by $ [...]$ (\ref{[]} ).
Further 3-point amplitudes are obtained from the above by cyclicity, the
remaining ones vanish. All cases can be summarized as
\be \label{M3h}
 M(h_1,h_2,h_2) = <12>^{2\eta h_3 +1 } <23>^{2\eta h_1 +1 } <31>^{2\eta
h_2 +1 }, \ \ \ h_1+h_2+h_3 = \eta, \eta = \pm 1.
\ee

\subsection{The Bjorken asymptotics}

Deep-inelastic electron - proton scattering  $e P \to e X $ ($X$ denotes
multi-hadron states ) is the prominent example of processes described by the
Bjorken asymptotics of QCD. The electron scatters at large momentum transfer $q$,
$-q^2 = Q^2 \gg M^2 $. 
The data on this process allow to resolve the structure of the hadron $P$ 
into quasi-free quarks and gluons \cite{BjP69}. 
This heuristic view is confirmed by the analysis of the
relevant Feynman graphs in the leading $\ln Q^2$  approximation
\cite{GL72a,GL72b, LNL74}. 

The inclusive cross section factorizes  into the
leptonic and hadronic tensor factors. 
The latter 
 can be considered as the s-channel
discontinuity of the virtual Compton scattering amplitude, $\gamma^* P \to
\gamma^* P$.  It expands into Lorentz structures multiplied by structure
functions $F(z, Q^2), z= \frac{Q^2}{2 (pq)}$. 

\begin{figure}
 \centering
 \includegraphics[width=0.4\linewidth]{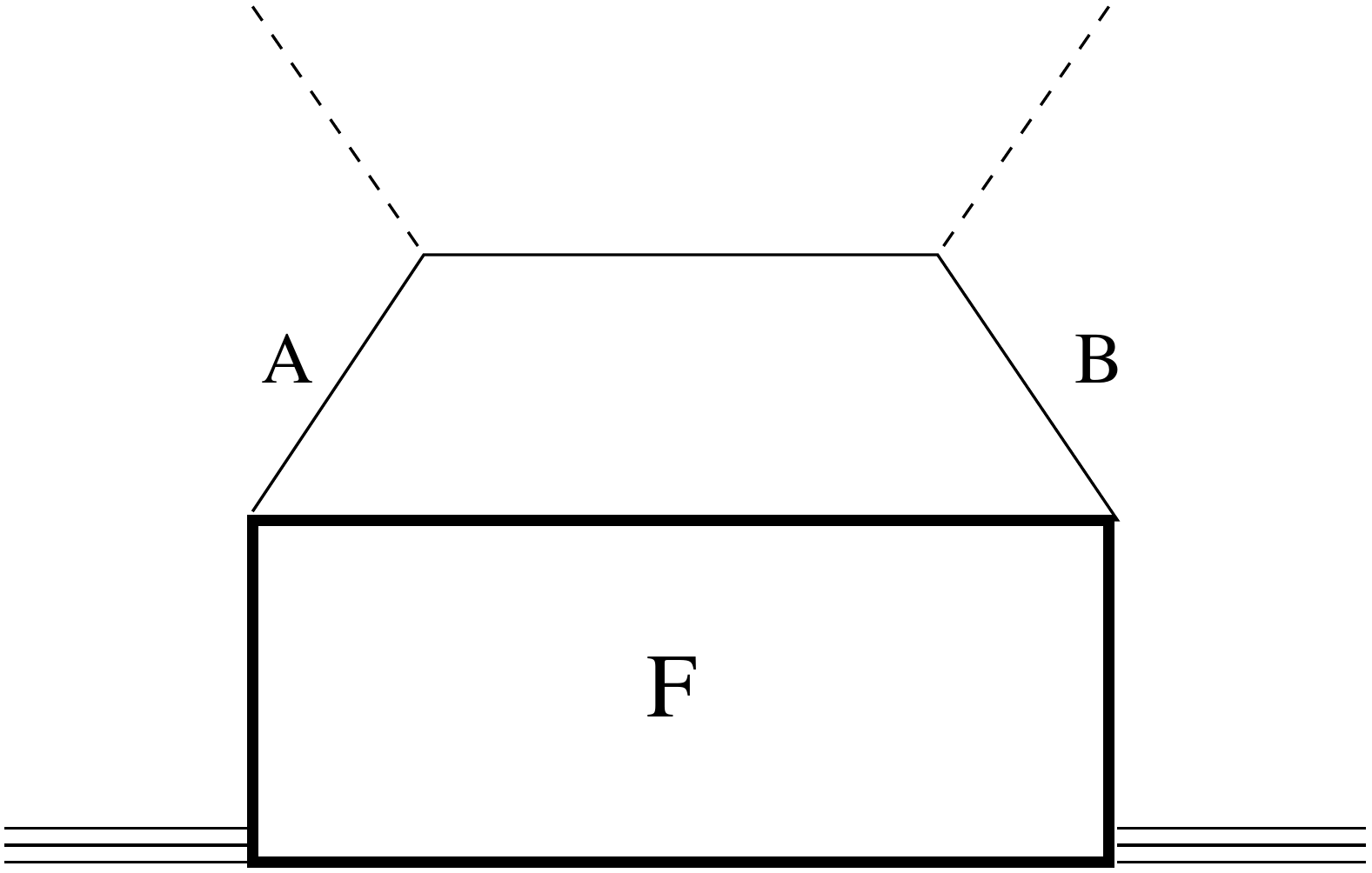}
 \caption[Virtual Compton scattering.]{Virtual Compton scattering, contribution of the structure function F.}
 \label{fig:fab}
 \end{figure}

 In the  perturbative expansion  of the hadronic tensor 
 we see (Fig. \ref{fig:fab} )  the virtual Compton scattering off a quark
convoluted with the structure function of the hadron $F(z, Q^2)$.
 The Bjorken variable $z = \frac{Q^2}{2 (pq)}$ is the  fraction of
the proton momentum $p$ carried by the  quark and $F(z, Q^2)$ can be
interpreted as the probability
distribution for finding in the target 
the particular parton with momentum fraction $z$.
 
 The asymptotics is governed by the
Feynman graphs contributing a factor $\ln Q^2 $ with each loop, i.e. the leading
logarithmic approximation applies.
 In the light cone gauge the
leading graphs simplify to ladders with self-energy and vertex corrections
inserted. The leading contribution arises from the integration region of the 
loop momenta $k_i = k^-_i n_- + k^+_i n_+ + \underline \kappa_i $ of
strongly ordered transverse components
\be \label{Bjorder}
 Q^2 \gg \kappa_N^2 \gg ... \gg \kappa_1^2 \gg M^2 . \ee
The loop integral reduces to the one over $k^+_i $, proportional to the
momentum fraction of the proton, and
the integration over the
remaining components reduces to $ \int \frac{d^2 \kappa}{\kappa^2} $ to
produce the required factor $\ln Q^2$.

The ladder rungs represent the partons hadronizing to the
final hadronic state X. The sum of ladder contributions to the structure
function leads to the equation  for
the structure function \cite{GL72a,YLD77,AP77}. 

The elementary building block of the ladder is reduced in the limit
(\ref{Bjorder}) to
 the 4-point function $W_{ABCD}$ depending on the $k^+$ momentum components
only, $k^+ = \frac{2}{\sqrt s} z$. 
The sum of the leading ladder contributions
including the  self-energy and vertex corrections leads to the equation for
the scale dependence of the structure function. 
$ A, B, C, D$ denote the parton type and helicity ($\pm 1 $ for gluons, $\pm
\half$ for quarks or anti-quarks). In the case of deep inelastic scattering
$B$ is related to $A$ and also $C$ to $D$ by conjugation,
$W_{A, \bar A; C , \bar C} \to W_{A; C}$. 

\be \label{branch}
\frac{d}{d\xi}  F_{A }(z; Q^2) = - W_A F_{A }(z; Q^2) +
\sum_{C} \int_z^1 \frac{dz\p}{z\p}
W_{C,A}(\frac{z}{z\p} ) F_{C } (z\p, Q^2). \ee
The scale dependence is formulated in terms of 
$ \xi (Q^2) = - \frac{2N_C}{ b} \ln \frac{\alpha(Q^2)}{\alpha(M^2)}, $
depending on the one-loop QCD coupling 
 $ \alpha(Q^2) = \frac{\alpha(M^2)}{1+ b \frac{\alpha(M^2)}{2\pi} \ln
\frac{Q^2}{M^2} }  , b = \frac{11}{3} N_C - \frac{2}{3} N_F $.
$N_c$ stands for the number of colors and $N_F$ for the number of quark
flavors .  

This form is the one of a branching process with $W_{A, B}(z)$
being the rate of the parton $A$ to decay into a parton $B$ carrying the
momentum fraction $z$ of $A$ and
$$ W_A = \sum_B \int_0^1 dz z W_{BA}(z). $$ 
The structure function  $F_{A }(z; Q^2) $ is proportional 
to the density of partons $A$ in the target hadron. 
Actually $W_{A, B}(z)$ is singular at $z=1$ and correspondingly $W_A$, 
and this branching picture
becomes meaningful with a regularization at $z=1$ only.

With the definition of the evolution kernel as the well-defined distribution
\be \label{Pkernel} P_{BA}(z) = W_{B,A}(z) - \delta_{A,B} \delta(1-z) W_A \ee
we have the Altarelli-Parisi form \cite{AP77},
\be \label{AP} \frac{d}{d\xi}  F_{A }(z; Q^2) =\sum_{C} \int_z^1
\frac{dz\p}{z\p}
P_{C,A}(\frac{z}{z\p} ) F_{C } (z\p, Q^2). \ee
It describes the scale $Q^2$ dependence of the structure functions and is
usually referred to as DGLAP evolution equation.

The spectra of the operators defined by such kernels 
are obtained by calculating the  moments
\be \label{anom} \gamma_{AB}(n) = \int_0^1 dz z^{n-1} P_{AB}(z) \ee
and diagonalizing  the resulting matrices in the parton type indices $A, B$. 

The hadronic tensor factor can be formulated also as the
hadron states matrix element  of the product of electromagnetic currents,
\be \label{PjjP} \int d^4 x e^{iqx} < P\p | T j_{\mu}(0) j_{\nu}(x) |P>. \ee
$T$ stands for time-ordering.
The treatment of the scale dependence can be formulated in terms of the
operator product expansion and renormalization group \cite{ChHM72, P74}.

In the case at hand both hadron states coincide, the general case appears in 
semi-exclusive hard processes.
 In the general case in the Lorentz expansion of the hadronic tensor factor
more structures contribute and the $n_+$ component of the momentum
transfer appears as an additional variable in the structure functions,
called generalized parton distributions \cite{ChZh77,ER80,BL79,Ji04}.

We see in both formulations of the hadronic tensor factor described
 that at the asymptotics the dynamics becomes one-dimensional. 
In the operator product form (\ref{PjjP}) the dominant
integration region is  $x^2 \sim \frac{1}{Q^2}$. The 
leading contributions in the operator product expansion are  products of two
light cone components of quarks ($f^*(x), f(x) $) or gluons ($A^*, A$)
located at the light ray. Let us write such quasi-parton operator
contributions as
\be \label{bilocal} F_{A B}(x_1, x_2; Q^2) = < P\p | f^*_A(x_1) f_B(x_2) |
P>,\ee
where$ A, B$ denote the parton type and helicity ($\pm 1 $ for gluons, $\pm
\half$ for quarks or anti-quarks) and $x_1, x_2$ are points on the light
ray. The angular brackets stand for hadronic states. 

For the field modes producing the leading contributions 
the QCD light cone action reduces to an one-dimensional effective action.
In particular from the kinetic terms (\ref{S2}) we get schematically
\be \label{S2Bj}
 S_{2 Bj} = \int d^1 x \mathcal{L}_{2 Bj}, \ \ 
\mathcal{L}_{2 Bj } = 2 A^* A + i f^* \dd^{-1} f.   \ee 
The scaling dimensions of the effective parton fields are
$\ell = - \half $ for gluons and $\ell = -1 $ for quarks.

By renormalization group the operator product contributions 
depend on momentum scale as
\be \label{ERBL} \frac{d}{d\xi}  F_{A B}(x_1, x_2; Q^2) = - \half (W_A + W_B)  \delta_{A C}
 \delta_{B D} \delta(x_1-x_1\p) \delta(x_2-x_2\p) + \ee $$   
\sum_{CD} \int dx_1\p dx_2\p W_{CD,AB}(x_1\p,x_2\p; x_1,x_2)  
F_{C D} (x_1\p,x_2\p; Q^2) $$
 
The  evolution equation (\ref{branch}) 
is the light-cone momentum representation of the latter equation
restricted to the case of vanishing momentum transfer. The position form
is well suited for the case  of generalized parton distributions.
The eigenvalues of the operators defined by the kernels in light-cone position
(\ref{ERBL})
and momentum representation (\ref{branch}), (\ref{AP}) coincide. 
The eigenvalues are the anomalous scaling dimensions of
the corresponding linear combinations of local operator products 
obtained by expansion of the bi-local operators of (\ref{bilocal}) 
in powers of the positions $x_1, x_2$.

The dynamics reduced in the Bjorken asymptotics to one dimension 
preserves the conformal symmetry at the tree level. 
The only relevant momentum component  $k^+$ can be written as factorised into
one-component variables, in analogy to the two-component spinor 
representation of light-like Lorentz  vectors, $k^+ = \lambda \bar \lambda$.
 The second spinor components (\ref{lambdak}) vanish with the transverse
momenta in the asymptotics, the collinear limit. 

The basic QCD three-point amplitudes (\ref{M3h}) result in this limit in
the parton splitting amplitudes $M_{{ split}}$ \cite{KS17}. 
They lead to the kernels 
$W_{AB}$ in analogy to the
construction of the scattering amplitudes from the basic three-point
amplitudes. 

 Consider the spinors (\ref{lambdak}) in terms of light-cone momentum components
(\ref{Sudak}), substitute $k^+ = \frac{\sqrt s}{2} z $ and choose 
$ z_1 = 1, \kappa_1 = 0, z_2 = z , \kappa_2 = \kappa, z_3 = 1-z, \kappa_3 =
- \kappa $. The parton splitting amplitude $1 \to 2 + 3 $ is then
$$ \lim_{\kappa \to 0} \frac{1}{<23>} M(h_1,h_2,h_3) =
\frac{z^{\half} (1-z)^{\half} }{z^{\eta h_2} (1-z)^{\eta h_2} }. $$
We obtain the symmetric from of the parton splitting amplitude,
$z_i = \frac{2}{\sqrt s} k^+_i, \sum z_1+z_2+z_3 = 0$, $h_1+h_2+h_3 = \eta$,  
\be \label{M3split}
M_{{ split}}(z_1, z_2,z_3;h_1,h_2,h_3) =  z_1^{ \half-\eta h_1}  
z_2^{+ \half -\eta h_2} z_3^{\half -\eta h_3}. \ee
The splitting  amplitude results in the kernels of the above scale evolution
equation as (changing here from the convention of ingoing momenta to the
normal physical one)
\be \label{WMsplit} W_{  AB}(z) = \sum_C W_{ A BC}(z), \ \ W_{ ABC}(z)=
| M_{split} (1, -z,  z-1,h_A, -h_B -h_C) |^2.
\ee
We describe the action of the conformal symmetry in the Bjorken limit.
Consider as generators $M_{ab} = x_a\dd_b$ (\ref{Mab}), but now with the indices
restricted to $a,b,=1,2$ and also the analogy of the 
elementary canonical transformation (\ref{cantrans1}), (\ref{cantrans2}) to
the helicity form . Now $\alpha, \alphadot $  take only one value
so that the indices will be omitted.    We have here (compare (\ref{2hx}) )  
$$ 2\hat h = (\underline x \underline \dd) +1, \ \ (\underline x \underline \dd) = x_a \dd_a. $$  

Let the generators,  $ M_{ab} = x_a \dd_b $, 
act on functions of $x_a$. This representation is reducible to
homogeneous functions  of degree $2\ell$, 
$ \phi(x) (x_2)^{2\ell}, \ \ x = \frac{x_1}{x_2} $.  
$2\ell$ is the eigenvalue of the dilatation $(\underline x \underline \dd)$, 
and is related to the helicity as
\be \label{2h2l1} 2 h = 2\ell  +1,  \ee
and the conformal symmetry generators act on $\phi(x)$ as
\be \label{Mabx} (M_{ab}) = 
\begin{pmatrix}
x\dd + \ell & \dd \\
- x( x\dd + 2\ell) & x\dd - \ell 
\end{pmatrix}.
\ee
For $2\ell = 0$ the M\"obius transformations on the line 
parametrized by $x$ are generated.

This ratio variable $x$ is the one to represent the
 light-ray positions appearing above in the operator products and
$k= \lambda \bar \lambda$ is to represent the light-cone momenta.

\subsection{The Regge asymptotics}

The Regge asymptotics of the scattering $A B \to A\p B\p $ is approached in the
kinematics $ s \gg t $, where $s= (p_A + p_B)^2, t= (p_A - p_{A\p})^2= q^2   $.
Tree amplitudes with charges and (physical) helicities of $A$ and $B$ coinciding with the
ones of
$A\p$ and $B\p$, correspondingly, have a leading   contribution behaving as $s$, where  gluons
are exchanged in the $t$ channel. Amplitudes with  $n_q$ quarks exchanged behave
asymptotically as $s^{1-  \half n_q}$. 

\begin{figure}
 \centering
 \includegraphics[width=0.2\linewidth]{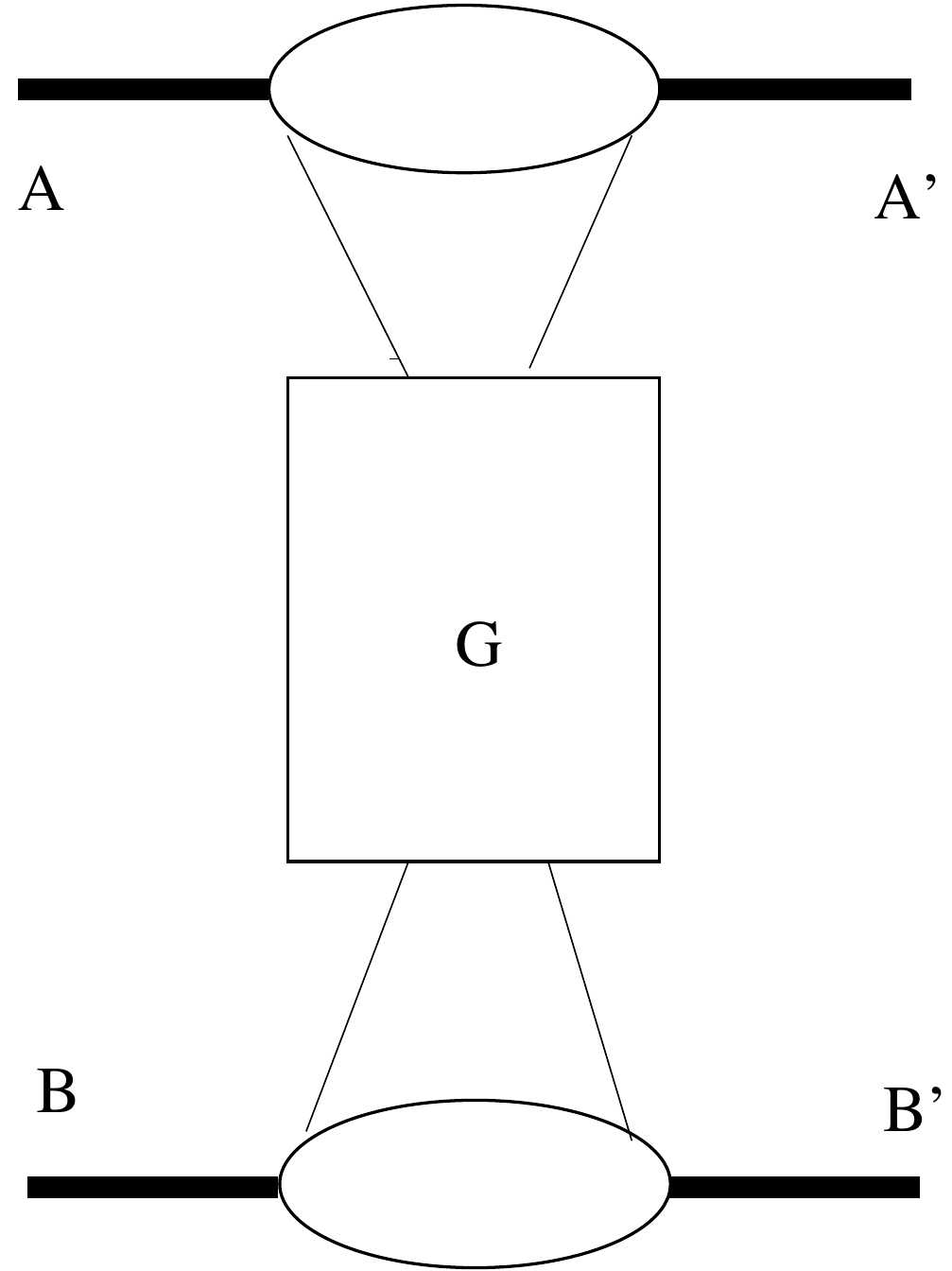}
 \caption[Mellin amplitude.]{Mellin amplitude. }
 \label{fig:impact}
 \end{figure}

It is convenient to consider the Mellin transform of the amplitude,
\be \label{Mellin} 
M(s, t) = \int_{- i\infty + \delta}^{+i \infty + \delta}
 d\omega F(\omega, q ) \left ( \frac{s}{M^2} \right )^{1+ \omega}. 
\ee
Here the details about signature are suppressed.
The Mellin amplitude $F(\omega, q)$, also called partial wave, appears as a convolution 
by integration over transverse momenta $\kappa_i$ of  
impact factors $\Phi_{A}(\underline \kappa) $, coupling to the particles     
$A$ and $A\p$, and $\Phi_{B}(\underline \kappa\p, )$, coupling to the particles
$B$ and $B\p$, with the reggeon Green function
$G(\omega, \underline\kappa,\underline \kappa\p) $, illustrated schematically
in Fig. \ref{fig:impact},
\be \label{impact}
F(\omega, q ) = \int d\underline{\kappa} d\underline{\kappa}\p 
\Phi_{A}(\underline{\kappa}) G(\omega, \underline{\kappa}, \underline{\kappa}\p) 
\Phi_{B}(\underline{\kappa}\p) \ee
Actually there is a sum over contributions of multiple reggeon exchanges
and $\underline{\kappa}$ stands for the set of reggeon transverse momenta.
We shall use also the convolution with one of the impact factors,  in the case
of two reggeons
$$ f(\omega, \kappa_1, \kappa_2) = \int d\kappa_1\p d\kappa_2\p
G(\omega, \kappa_1, \kappa_2, \kappa_1\p, \kappa_2\p)
\Phi_{B}(\kappa_1\p, \kappa_2\p). $$
 
The asymptotics in $s$ is determined by the right-most singularities in
$\omega$ of the reggeon Green functions $G(\omega,\underline{\kappa},
\underline{\kappa}\p)$ . 

The reggeon Green functions corresponding to the single gluon or quark exchange
are, correspondingly, 
\be \label{rgreen0} 
 \frac{1}{|\kappa |^2} \frac{1}{\omega}, \ \ \ \
 \frac{1}{\kappa } \frac{1}{\omega + \half},  \ee
or the  complex conjugate for the opposite quark helicity.
The transverse momentum dependence can be described by the propagators of
the effective reggeon fields in two transverse dimensions,
\be \label{RS2} S_{2 R} = \int d^2 x \mathcal{L}_{2 R}(x), \ \ 
\mathcal{L}_{2 R} (x) = - \mathcal{A}_+(x) \dd \dd^* \mathcal{A}_-(x) -
\chi_+(x) \dd \chi_-(x) - \tilde \chi_+(x) \dd \tilde \chi_-(x). \ee

The reggeon fields with subcript $+$ are to couple to the impact factor
$\Phi_A$ and to particles flying almost in the direction $p_A$ and 
the fields with subscript $-$ are to couple to particles flying almost
in the direction  $p_B$. 

In the leading logarithmic approximation the conformal symmetry is
preserved, acting by linear fractions (M\"obius transformations)
on the transverse positions, separately for $x$ and $x^*$. 
We have in particular scale symmetry $ x \to \lambda x, x^*
\to \bar \lambda x^* $ for the corresponding transformation of the fields
as appearing in the kinetic terms (\ref{RS2}) 
with the scale exponents $(\ell,\bar\ell)$ being $(0,0)$ for gluons and 
$(0, -\half)$ or $(-\half,0)$ for quarks.

In the leading logarithmic approximation the Feynman graph contributions
proportional to $\ln s$ for each loop are summed. In the light-cone gauge 
the leading graphs simplify to  ladders with the 
t-channel exchanged gluons or quarks
being reggeized by loop corrections. The kinematics of the leading ladder
contributions are characterized by the strong ordering conditions
on the loop momentum components $k^+, k^-$,
\be \label{Rordering} k^+_1 \gg ...\gg k^+_N , \ \ \   k^-_1 \ll ...\ll
k^-_N.
\ee
The integration over these components reduces to the logarithmic form
$\sim \int \frac{dk^+}{k^+} $ and, consequently, 
the dynamics reduces to the two transverse dimensions. 

\begin{figure}
\centering
\includegraphics[width=0.7\linewidth]{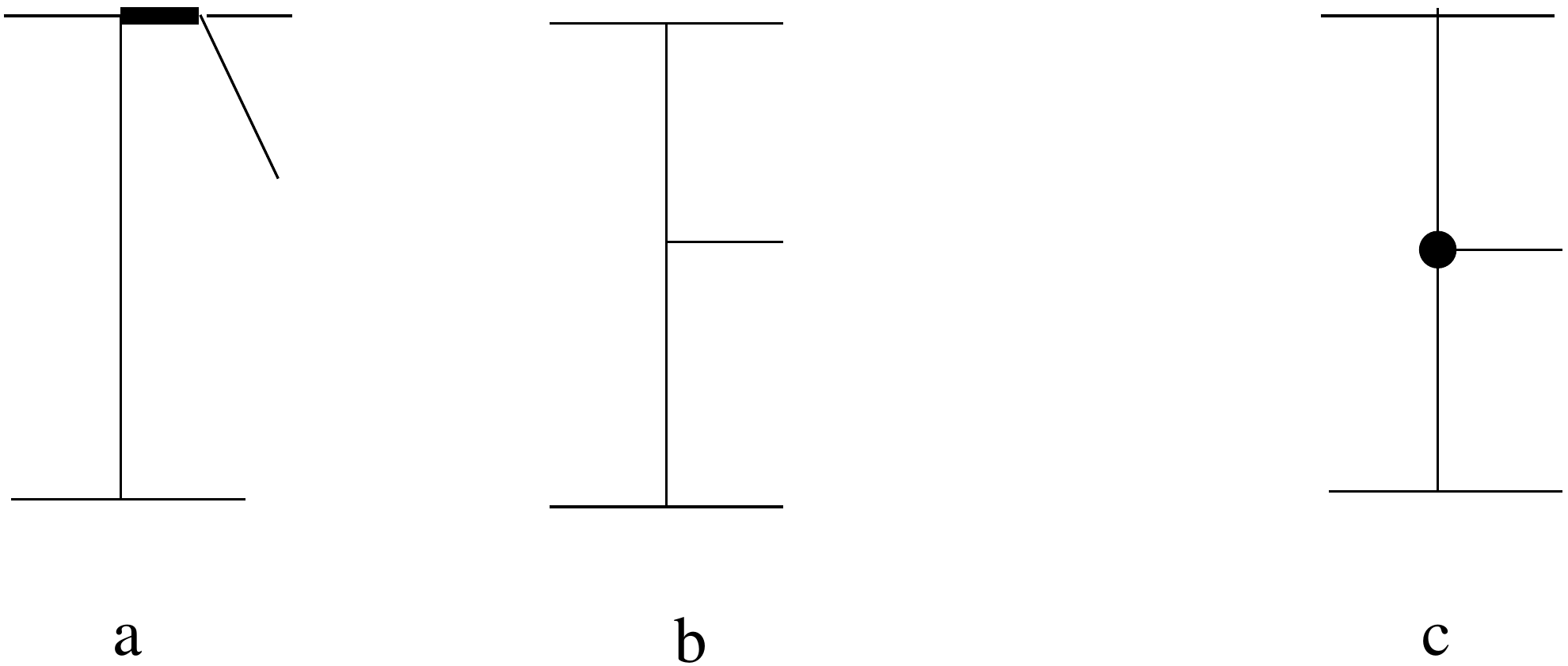}
\caption[Contibutions to $2 \to 3$ scattering.]{$2 \to 3$ scattering. a) Bremsstrahlung, 
b) Direct emission, c) Effective production vertex.}
\label{fig:prod}
\end{figure}

The elementary building block of the ladder and the reggeization can be
understood in terms of the effective production vertex \cite{KLS94}. 
Consider the 
scattering of two partons with production of a parton. The sum of the bremsstrahlung 
off the scattered parton Fig. \ref{fig:prod}a and the emission from the exchanged
reggeon Fig. \ref{fig:prod}b is represented by the emission with the effective 
production vertex $V_{gg}$ Fig. \ref{fig:prod}c.

 For the gluonic reggeons we have in transverse momenta written in
complex numbers the effective production vertex
\be \label{Vgg}
 V_{gg} ( \kappa, \kappa\p) = \frac{\kappa^* \kappa\p}{\kappa- \kappa\p}
\ee 
or as  an effective action term  to be added to (\ref{RS2}) 
$$\mathcal{L}_p (x) = g (\dd \mathcal{A}_+(x)  \dd^* \mathcal{A}_- (x))
a(x).
 $$
$a(x)$ stands for the emitted gluon. Its momentum range is close to the mass
shell.  $\mathcal{A}_{\pm} $ are the exchanged  gluonic reggeons with
momenta in the range $k^+ k^- \ll \kappa \kappa^* $. 

Instead of the argument in terms of graphs Fig. \ref{fig:prod}
 the effective production term can be obtained by integrating out
(elimination by equation of motion) the modes far off shell,
$k^+ k^- \gg \kappa \kappa^* $, represented by the fat line in 
Fig. \ref{fig:prod}a.

\begin{figure}
\centering
\includegraphics[width=0.7\linewidth]{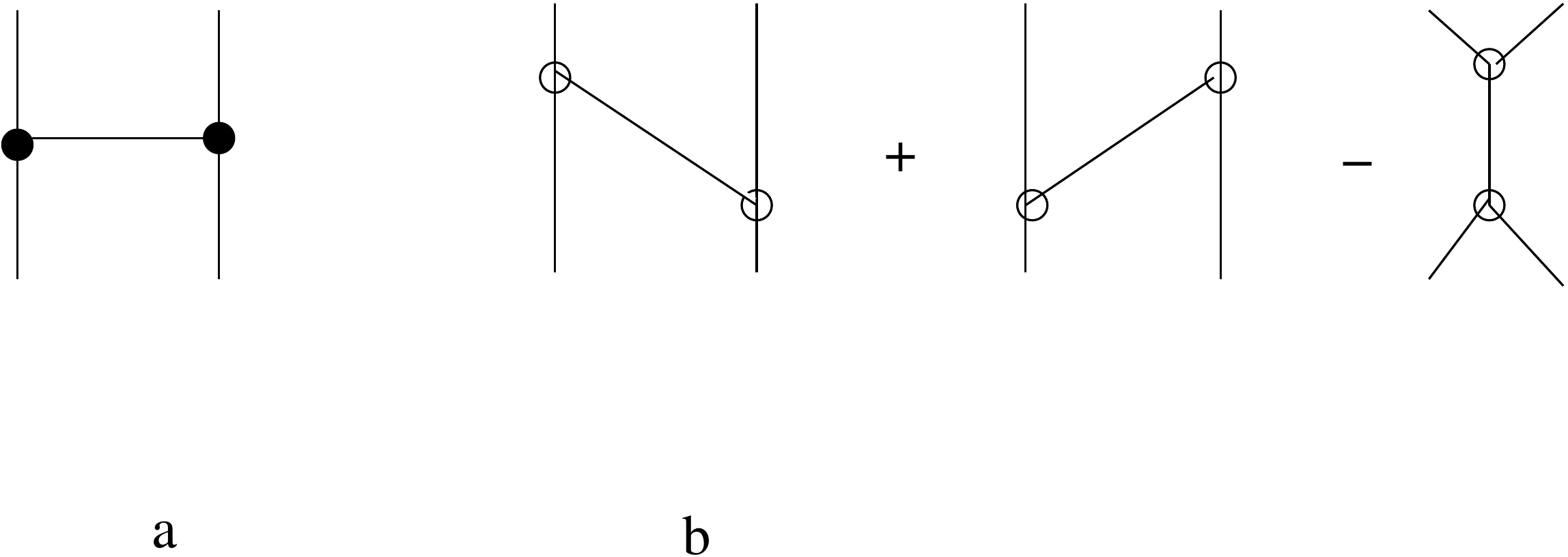}
\caption[Elementary ladder block.]{Elementary ladder block, 
composed  by effective production vertices (a) or by  reggeon triple vertices
(b).}
\label{fig:block}
\end{figure}

The contribution of the elementary building block of the ladder 
Fig. \ref{fig:block}a
is obtained from this effective production vertex (black dot in the figure) as
\be \label{K0}
 K^{(0)}(\kappa_1, \kappa_2, \kappa_1\p, \kappa_2\p ) =
V_{gg}( \kappa_1, \kappa\p_1) V^*_{gg}( \kappa_2, \kappa\p_2) + c.c. \ee
The one-loop contribution to the gluon reggeisation can be written as  
 \be  \label{traj}\omega_g(\kappa) = \frac{1}{2 (2\pi)^3} |\kappa|^2 \int
\frac{d^2 \kappa\p}{|\kappa\p|^2 |\kappa - \kappa\p|^2}.  \ee

Including this contribution modifies the Green function of the reggeized
gluon from (\ref{rgreen0}) to 
\be \label{rgreen} 
\frac{1}{|\kappa|^2} \  \frac{1}{\omega + g_r^2  \omega_g(\kappa) }, \ \ \ g_r^2 = 
\frac{g^2 N_c}{2 (2\pi)^3}. \ee
The Green function of two non-interacting
reggeised gluons  is
$$ G^{(0)}(\omega, \kappa_1, \kappa_2, \kappa_1\p, \kappa_2\p)
= |\kappa_1|^{-2} |\kappa_2|^{-2} \delta(\kappa_1- \kappa_1\p) \delta(\kappa_2-
\kappa_2\p) f^{(0)}(\omega, \kappa_1, \kappa_2), $$
$$  f^{(0)}(\omega, \kappa_1, \kappa_2) = 
\frac{1}{\omega + g_r^2 \omega_g(\kappa_1) +  g_r^2 \omega_g(\kappa_2) }
$$

The sum of the ladder graph contributions of the two-reggeon interaction
is obtained by solving  the equation
$$  G(\omega, \kappa_1, \kappa_2, \kappa_1\p, \kappa_2\p) =   
G^{(0)}(\omega, \kappa_1, \kappa_2, \kappa_1\p, \kappa_2\p)
+ g_r^2 \int \frac{d^2\kappa_1\p
d^2\kappa_2\pp}{|\kappa_1\pp|^2 |\kappa_2\pp|^2 }  
K^{(0)}(\kappa_1\pp, \kappa_2\pp, \kappa_1, \kappa_2 )
 G(\omega, \kappa_1\pp, \kappa_2\pp, \kappa_1\p, \kappa_2\p). $$
The reggeization contribution $\omega_g(\kappa)$ (\ref{traj})
becomes well defined with
regularization only and the  bare kernel $K^{(0)}$ defines an operator
only with regularization. We rewrite the equation in terms of the BFKL
kernel resulting in a well defined operator at vanishing regularization
parameter \cite{L76, FKL, BL78},
$$  K(\kappa_1\p, \kappa_2\p, \kappa_1, \kappa_2 ) =
 K^{(0)}(\kappa_1\p, \kappa_2\p, \kappa_1, \kappa_2 ) - 
(\omega_g(\kappa_1) + \omega_g(\kappa_2) )  \delta(\kappa_1- \kappa_1\p) \delta(\kappa_2-
\kappa_2\p),
$$
\be \label{BFKL}  \omega f(\omega, \kappa_1, \kappa_2) = 
g_r^2 \int \frac{d^2\kappa_1\p
d^2\kappa_2\p}{|\kappa_1\p|^2 |\kappa_2\p|^2 }  
K(\kappa_1\p, \kappa_2\p, \kappa_1, \kappa_2 )
f(\omega, \kappa_1\p,\kappa_2\p). \ee

The kernel can be reconstructed from elementary  reggeon triple vertices
\cite{RK95},
obtained from the transverse part of the induced vertex obtained from
the bremsstrahlung contribution Fig. \ref{fig:prod}a.
$\mathcal{V}^{+--}(\kappa_1, \kappa_2) $ is the interaction vertex of a reggeon $\mathcal{A}_+$
with two reggeons $\mathcal{A}_-$, $\kappa_1, \kappa_2$ are the momenta of
the latter.    
\be \label{V-++}
 \mathcal{V}^{-++} (\kappa_1, \kappa_2) = 
\mathcal{V}^{+--} (\kappa_1, \kappa_2) = |\kappa_1 + \kappa_2|^2 . \ee
 
\begin{figure}
\centering
\includegraphics[width=0.05\linewidth]{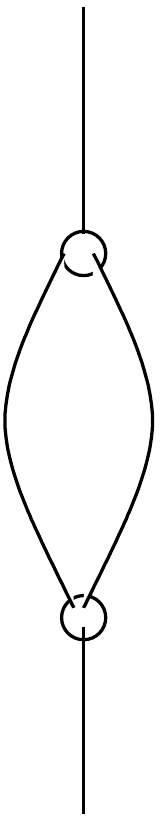}
\caption[Reggeization loop.]{Reggeization loop composed of reggeon triple vertices.}
\label{fig:traj}
\end{figure}

 Fig. \ref{fig:block}b  shows the graphs with these vertices 
resulting in the bare BFKL kernel $K^{(0)}$
(\ref{K0})),
standing term by term for
\be \label{K0V}
 \frac{|\kappa_2|^2 |\kappa_1\p|^2 }{|\kappa_1-\kappa_1\p|^2} +
\frac{|\kappa_1|^2 |\kappa_2\p|^2 }{|\kappa_1-\kappa_1\p|^2} - 
|\kappa_1 + \kappa_2|^2= K^{(0)} (\kappa_1\p, \kappa_2\p, \kappa_1, \kappa_2), 
\ee
 and Fig. \ref{fig:traj} shows the loop resulting in the reggeization $\omega_g$ (\ref{traj}).  
In Fig. \ref{fig:block}b and Fig. \ref{fig:traj} the edges of a 
triple reggeon vertex (open circle) pointing upwards
represent $\mathcal{A}_-$ and the edges pointing downwards represent
$\mathcal{A}_+$. 

The conformal symmetry of the equation (\ref{BFKL})
becomes evident  after the transformation to transverse positions. 
The Fourier transformation is to be done with care because of divergencies.
Schematically the result 
expressed as  an 
operator in terms of the canonical pairs $x_I, \dd_I, I = 1,2 $ and their complex
conjugates $x^*_I, \dd^*_I $
can be understood applying to the expression of the BFKL kernel
(\ref{BFKL})
the substitution rules
\be \label{rules}
 |\kappa_1-\kappa_1\p| \to - \ln |x_{12}|^2, \omega (\kappa_1) \to \psi(1)
- \ln (\dd_1 \dd_1^*), \ \ 
 \kappa_1 \to \dd_1^*, \ \kappa_2^* \to \dd_2. \ee
This leads to the equation for the Fourier transform 
$\tilde f(\omega, x_1,x_2)$ 
\be  \label{BFKLx}
 \omega \tilde f(\omega, x_1,x_2) = \frac{g^2 N_c}{8 \pi^2} \hat H_g \  
 \tilde f(\omega, x_1,x_2) \ee
with the BFKL hamiltonian $ \hat H_g$ decomposing into  a holomorphic and an 
anti-holomrophic parts, the first involving $x_I, \dd_I, I = 1,2 $ only and the second 
$x^*_I, \dd^*_I, I = 1,2 $ only.
$$ \hat H_g = (H_g + H^*_g), $$ 
\be \label{Hg1} H_g = 2\psi(1) - \dd_1^{-1} \ln x_{12} \ \dd_1 -  \dd_2^{-1} \ln x_{12}
\ \dd_2 - \ln \dd_1 - \ln \dd_2 .\ee
Further formal operator manipulations lead to the expression 
in terms of the tensor Casimir operator 
\be \label{Casimir} C_{12} = (M^{(1)}+M^{(2)})_{ab} (M^{(1)}+ M^{(2)})_{ba},
\ee  
where $M^{(I)}_{ab}$ are the operators in the matrix elements of (\ref{Mabx})
with $x, \dd$ replaced by $x_I, \dd_I$ and $\ell$ by $\ell_I$. 
The symmetry algebra consists of two copies of $g\ell(2)$, generated by
 $M_{ab}$ acting on $x$ but not on $x^*$ and $M^*_{ab}$ acting on $x^*$. 
Correspondingly we have two labels $(\ell, \bar \ell)$ for the representations,
$\ell$ referring to the first and $\bar \ell$ referring to the second.

In the case of two gluonic reggeons considered  the representation
parameters are $(\ell_1,\bar \ell_1),  (\ell_2, \bar \ell_2)= (0, 0), (0, 0)$. 
 This corresponds to the above
remark  after (\ref{RS2}) about the scale dimensions of the reggeon fields. 
The formulation
generalizes to more than two reggeons and to the inclusions of fermionic
reggeons. The latter are described by the representations $(-\half, 0)$ or 
$(0, -\half)$ depending on helicity and chirality \cite{RK05}. 

The colour-dipole model \cite{NNN94,AHM94,YB99,YK99} 
of high-energy scattering leads to the
integral form of the Hamiltonian
\be \label{dipole} \hat H_g \tilde f(\omega, x_1,x_2) =
\frac{1}{\pi} \int d^2x_0 \frac{|x_{12} |^2 }{|x_{10}|^2 |x_{20}|^2 }
\left ( \tilde f(\omega, x_1,x_2) - \tilde f(\omega, x_1,x_0) - \tilde f(\omega,
x_0,x_2) \right ).  \ee
$\tilde f(\omega, x_1,x_2)$ can be regarded as the Mellin transform of the
dipole distribution with respect to their end-points in the transverse plane 
and the equation (\ref{BFKLx}) describes its evolution  with  rapidity $Y \sim \ln s$.  

We present the arguments of \cite{BLV04} leading to this form.
By the formal operator relations
\be \label{rel} 
\dd \ln x \ \dd^{-1} = \ln x + x^{-1} \dd^{-1}, \ \ x \ln \dd \ x^{-1} =
\ln \dd - \dd^{-1} x^{-1}  \ee
the Hamiltonian (\ref{Hg1}) is rewritten as
\be \label{Hg2} 
\hat H_g =  4 \psi(1) - 2 \ln |x_{12}|^2 - 
2  |x_{12}|^{-2} \ln |\dd_1 \dd_2|^2   |x_{12}|^2 .
\ee
 Regularizing the kernel in (\ref{dipole}) we have  
$$ \int \frac{d^2 x_0}{\pi} \frac{|x_{12} |^2 }{(|x_{10}|^2+ \delta^2) (|x_{20}|^2+
\delta^2) } = \int_0^1 d \alpha  
\frac{|x_{12} |^2 }{(\alpha (1-\alpha) |x_{12}|^2+ \delta^2} 
= 2 \ln {\frac{|x_{12}|^2}{\delta^2}} + \mathcal{O}(\delta^2) $$
$$ - \int \frac{d^2 x_0}{\pi} \frac{|x_{12} |^2 }{(|x_{10}|^2+ \delta^2) (|x_{20}|^2+
\delta^2) } \tilde f(\omega, x_1, x_0) = 
|x_{12} |^2 ( \ln(\delta^2 |dd_2|^2 - 2\psi(1) ) |x_{12} |^{-2} \tilde f(\omega, x_1,x_2)
+ \mathcal{O}(\delta^2 ) $$
This shows that the integral operator (\ref{dipole}) acts in the limit of vanishing
$\delta$ as the operator (\ref{Hg2}). 

The eigenvalues of the operator defined by the kernel $K (\kappa_1\p,
\kappa_2\p, \kappa_1 \kappa_2) $ (\ref{BFKL})
are easily calculated in the transverse momentum form in the
forward limit, $\kappa_1+\kappa_2 = -\kappa_1\p-\kappa_2\p = 0$.
In this case the eigenfunctions are 
$$ E^{(n, \nu)} (\kappa) = |\kappa|^{ \half + i\nu} \left (
\frac{\kappa^*}{|\kappa|} \right )^{n}, $$
a basis of functions  on the complex plane for
integer $n$ and $\nu \in \R^1$. The eigenvalues can be written as
\be \label{Omegag}
\Omega_g(n, \nu) = \chi_0(m) + \chi_0(\bar m), 
\chi_0(m) = 2 \psi(1) - \psi(m) - \psi(1-m), \ee
$$ m= \half + i\nu + \half n, \ \  \bar m = \half + i\nu - \half n . $$
$\chi_0(m)$ is actually a function of $m(1-m)$, the eigenvalue of the
Casimir operator (\ref{Casimir}). 
In the form of transverse positions the eigenfunctions are
the conformal three-point functions. 
\be \label{Ex} 
 E^{(n, \nu)}(x_1,x_2;x_0) = E_{m} (x_1,x_2;x_0) E_{\bar m}
(x^*_1,x^*_2;x^*_0), \ee 
$$ E_{m} (x_1,x_2;x_0) = \left ( \frac{x_{12} }{x_{10} x_{20}} \right )^m, $$
obeying the completeness relation  
$$\delta^{(2)}(x_1 - x_1\p ) \delta^{(2)}(x_2 - x_2\p ) = 
\int d^2x_0 \int_{-\infty}^{+\infty} d\nu 
\sum_{n=-\infty}^{+\infty} $$ $$
\frac {\nu^2+ \frac{n^2}{4} }{\pi^4 |x_{1 1\p}|^2 x_{2 2\p}|^2 } 
E^{(n, \nu) *} ( x_1, x_2, x_0) E^{(n, \nu) } ( x\p_1, x\p_2, x_0). 
$$
 The orthogonality relation expresses the fact that this basis is twofold
over-complete \cite{KL89}.  

The spectral properties are the basis for giving the above forms of 
the BFKL operator and the unconventional steps of transformations
(\ref{rules}), (\ref{rel}) well defined meanings.

The renormalized coupling, depending on the scale $Q^2$, enters the 
Bjorken limit evolution equations at leading order. In the BFKL equation,
describing the Regge asymptotics, the bare coupling enters at leading order.
Its renormalization is included at next-to-leading order only 
\cite{FL98},\cite{CC98}, \cite{KL00}.

For the region of overlap of both asymptotics combinations of both types of
equations have to be considered, including the BFKL contributions as
corrections to the DGLAP equation \cite{CCH90} or improving the BFKL
equation by the renormalization group \cite{KL89}. 
In the overlap region the contributions of muti-reggeon exchange become 
important. They can be described by non-linear terms to be included into the 
BFKL equation \cite{Bartels80, KP80}. In the dipole picture the multi-reggeon
contributions appear as creation of more dipoles \cite{YB99, YK99}.
In the ultimate asymptotics a regime of saturation emerges 
\cite{McLV93, GIJMV10}
allowing the application of thermodynamic concepts.
Characteristics of the thermodynamic limit of models directly
related to the integrable spin chains, where the BFKL hamiltonian
describes the pair interaction, have been proposed to be applied here
\cite{KhL17, KhKorepin21}.

\section{Yangian symmetry } 
\setcounter{equation}{0}

\subsection{Monodromy matrix operators and symmetric correlators}

We start with the formulation of the $g\ell(n)$ Lie algebra relations as the
fundamental Yang-Baxter relation
\be \label{fundYB}
 \mathcal{R}(u-v) (L(u)\otimes I) (I\otimes L(v)) = (I\otimes L(v))( L(u)\otimes I) 
\mathcal{R}(u-v).
\ee
Here $\mathcal{R}(u-v)$ stands for Yang's $R$ matrix. Its size is $n^2\times
n^2$
in our case and  composed of the unit matrix and the matrix representing the
permutation of the  factors in the tensor product of the
$n$-dimensional fundamental representation spaces,  
$ \mathcal{R}(u) = u I_{n^2\times n^2} + P $. The $L$ matrices are of size
$n \times n$ and linear in the
spectral parameter $u$, $L(u) = u I_{n\times n} +  L $.
 $L_{ab}, a,b, = 1, ...,n$
are the generators of the  Lie algebra $g\ell(n)$. 
In index notation we have
\be \label{LabR}
L_{ab}(u) = u \delta_{ab} + L_{ab}, \ \ \ 
\mathcal{R}^{a_1a_2}_{b_1b_2}(u) =  u \delta^{a_1}_{b_1} \delta^{a_2}_{b_2}
+ \delta^{a_1}_{b_2} \delta^{a_2}_{b_1}.
\ee  

The Yangian algebra can be introduced by the matrix operator $T(u)$
obeying the above Yang-Baxter  relation (\ref{fundYB}) with  $L(u)$  
substituted by $T(u)$ and where now
the dependence on $u$ is not restricted as above. Then the generators of the
extended Yangian algebra \cite{Drinfeld}, \cite{Molev} appear in the  expansion of $T(u)$ 
in inverse powers of $u$ and the extended
Yangian algebra relations  follow from the Yang-Baxter  relation
(\ref{fundYB}).

We shall deal with  evaluations  of the $g\ell(n)$ Yangian of finite order $N$
where $T(u)$ is constructed from the $L$ matrices in the way well know in 
the context of integrable models. 
Let the matrix elements of $L_I(u_I)$ act on the representation space $V_I$
and consider  the matrix product defining the monodromy matrix    
\be \label{monodromy}
T(u,\delta_1, \dots,\delta_N)= T(\mathbf{u}) = L_1(u_1) \cdots L_N(u_N), \quad
u_I = u+\delta_I,  \ee 
acting on the tensor product $ V_1 \otimes \cdots \otimes V_N$.

 We consider  the case of representations by  homogeneous
functions of  $\mathbf{x}_I = (x_{i,1},..., x_{I,n}) $  
and the weight $2\ell_I$ is the degree of homogeneity, 
\be \label{x2l}
 \psi(\mathbf{x}_I) = x_{i,n}^{2\ell_I} \phi  (\mathbf{x}\p_I), \ \ \  
\mathbf{x}_I\p = (\frac{x_{I,1}}{x_{I,n}},..., \frac{x_{I,n-1}}{x_{I,n}}).
\ee
The elements of the tensor product $ V_1 \otimes \cdots \otimes V_N$ are 
functions of $N$ points in the $n$ dimensional space.

We consider  the  Lie algebra generators of the
Jordan-Schwinger (JS) form, 
\be \label{L+-}
 L^+_{I, a b} = \dd_{I,a} x_{I,b}, \  \ \  
L^-_{I, a b} =  - x_{I, a} \dd_{I,b}, 
\ee
built from $n$ canonical pairs $x_{I,a} ,\dd_{I, a}$ at each point
$I$, 
\be \label{Heisenb} 
[\dd_{I, a}, x_{J, b} ]= \delta_{I J} \delta_{ab}. \ee
Notice that the elementary canonical transformation
\be \label{canon} \begin{pmatrix}
x_{I,a} \\ \dd_{I,a} 
\end{pmatrix} \rightarrow
 \begin{pmatrix}
\dd_{I,a} \\ -x_{I,a} 
\end{pmatrix} \ee
transforms $L_I^+(u)$ into $L_I^-(u)$. 

We have the matrix inversion relations (for any point $I$)
\be \label{inv}
 \left ( \frac{L^+ (u)}{u} \right )^{-1} = \frac{L^+(-u-1-
(\underline{x}\underline{\dd}))}{
-u-1- (\underline{x}\underline{\dd})},  \ \ \ 
\left ( \frac{L^- (u)}{u} \right )^{-1} = \frac{L^-(-u+1 +
(\underline{x}\underline{\dd}))}{
-u+1+ (\underline{x}\underline{\dd})},   
\ee
where $(\underline{x}\underline{\dd}) = \sum_{a=1}^n x_a \dd_a $. 

The two forms of JS generators are also related by matrix transposition,
$(L^t)_{ab} = L_{ba}$,
\be \label{Lt} (L^+(u) )^t = - L^-(-u-1). \ee
This transposition operation and the inversion act depending on ordering,
\be \label{Lt-1} (L^+(u))^{ t \ -1} = (L^+(u+n) )^{-1 \ t}. \ee

Related to integrals over the positions $x_a$  we consider another transposition
operation acting on the canonical pairs as
$$ (x_a)^T = x_a, \ \ \dd_a^T = - \dd_a, $$
 leaving indices unchanged but reversing the order in products of such
operators, similar to the matrix transposition, at the same point e.g.
$ (x_{I, a} \dd_{I, b} \dd_{J,c})^T =    (\dd_{I, b})^T  (x_{I,
a})^T  \dd_{J,c}^T $. 
\be \label{LT} (L^+(u) )^T = (L^-(u) )^t = - L^+(-u-1), \ \ \ 
 (\underline{x}\underline{\dd})^T = -n - (\underline{x}\underline{\dd}). \ee

We intend to consider representations  by functions of
$x_{I, a}$ homogeneous at each point $I$ (\ref{x2l}), i.e. we
restrict to the subspace of fixed eigenvalues of the dilatation operators,
$ (\underline{x}_I\underline{\dd}_I) \to 2 \ell_I $.
The corresponding projector will be denoted by $\Pi_I(2\ell_I)$.
 
The  projection of the $L$ operator depends on this degree of homogeneity,
\be \label{projL}
 \Pi_I(2\ell_I) L_I^+(u) \Pi_I(2\ell_I) = L_I^+(u_I^+, u_I), \ u_I^+ = u +
\delta_I + 2\ell_I \ee
The action of the inversion and the transpositions $t$ and $T$ on the
projected $L$ obey
\be \label{Lprojinv} ( L^+(u^+, u) )^{-1} = \frac{L^+(-u-1, -u^+ -1)}{u (-u^+ -1) },
\ee 
\be \label{Lprojt}
( L^+(u^+, u) )^{T} = - L^+( -u^+ -1-n, -u-1), \ \ \
 (L^+ (u^+, u) )^{t \ -1} = (L^+(u^+ +n, u+n) )^{-1 \ t}, \ee 
\be \label{LprojT-1}
(L^+ (u^+, u) )^{T \ -1} = \frac{L^+(u, u^+ +n)}{ (u+1) (u^+ +n)}, \ \ \
(L^+ (u^+, u) )^{-1 \ T} = \frac{L^+(u-n, u^+)}{u (u^+ +1)}.
 \ee

The projected $L$ obey the fundamental Yang-Baxter relation (\ref{fundYB}). 
In the following we prefer  the form $L^+$ and omit the superscript $+$. 

The fundamental Yang-Baxter relation (\ref{fundYB}) 
is fulfilled by the monodromy composed
as the ordered matrix product of $L_I (u_I)$ (\ref{monodromy}) as well as
 of the projected  $ L_I(u_I^+, u_I) $.
\be \label{projmono}
T(\mathbf{u}) = L_1(u_1^+, u_1) ...L_N(u_N^+, u_N), \ \ 
  \mathbf{u} =
\begin{pmatrix}
u_1 & ..., u_N \\
u_1^+ & ..., u^+_N 
\end{pmatrix}
\ee

$N$-point functions $\Phi $ homogeneous at each point obeying the condition
\be \label{YSC}
 T(\mathbf{u}) \ \Phi = I E(\mathbf{u}) \ \Phi \ee
are called Yangian symmetric correlators (YSC) \cite{CK}. 
On l.h.s an operator-valued matrix is acting and on r.h.s. there is the unit
matrix $I$ and  the eigenvalue  depending on the set of parameters
$\mathbf{u}$. The dependence of $\Phi$ on the parameters is invariant 
with respect to  shifts of the spectral parameter,
$ u_I \to u_I + v, u_I^+ \to u_I^+ + v$.
The YSC are invariants in the corresponding representation of the
Yangian algebra. 
  
 As solutions of (\ref{YSC}) we shall accept also expressions involving
distributions, extending the representation space $ V_1 \otimes \cdots \otimes
V_N$.  The $\delta$-distributions are to be understood in Dolbeault sense
(cf. \cite{GH}), their arguments  must not be restricted to real
values.

By elementary canonical transformations of the underlying canonical pairs
(\ref{canon}), where this transfomation may be applied to a subset of the 
index values $I, a$, the monodromies and the symmetric correlators 
acquire other representation forms.  In the applications to be considered
the helicity representation plays an important role. It will be desribed in
subsection 3.4. 

In the next subsection we outline methods of construction of YSC.
The Yang-Baxter operators and the corresponding $RLL$ relations,
being the working tool in one of them,
will be discussed in the subsection 3.3.

Examples of YSC constructions will be presented in subsection 3.5.
The role of YSC as kernels of symmetric integral operators 
will be considered in subsection 3.6.

\subsection{YSC construction methods}

We find trivial examples of YSC in the case of one point, $N=1$.

$$ L(u, u) \cdot 1 = (u+1) 1, 
\ \ \ L(u-n, u) \cdot \delta (\mathbf{x}) = u \delta^{(n)} (\mathbf{x}).
$$
The monodromy $T_{1,2.., N}$ can be considered as the product
$ T_{1,2.., N} = T_{1,2,..., N_1} T_{N_1+1, ..., N}$,   
and consequently a product of YSC,
\be \label{PhiPhi}  \Phi_{1, .., N_1} \cdot \Phi_{N_1, ..., N}, \ee
the first obeying (\ref{YSC}) with $T_{1,2,..., N_1}$ and the second
obeying (\ref{YSC}) with $T_{N_1+1, ..., N}$, is a YSC obeying 
(\ref{YSC}). 

\subsubsection{ The convolution method}

In the product correlator (\ref{PhiPhi}) the points with the labels $1, ..., N_1$
and $N_1+1, ..., N$ are not connected.
A way to get connected YSC from a product
goes by identifying some points of the first and the second factor and
integrating over these identified points. 
 We  illustrate this in the example of the convolution
of two 3-point correlators and derive the conditions on the
representation parameters for obeying the  YSC condition of the result.

We derive the conditions for a one-point convolution of YCS to be
YSC. Consider the convolution of two 3-point YSC
$$ \Phi_{1,2,3,4} = \int d \mathbf{x}_0 
\Phi_{1,2,0}(\mathbf{x}_1,\mathbf{x}_2,\mathbf{x}_0)
\bar  \Phi_{0,3,4}(\mathbf{x}_0,\mathbf{x}_3,\mathbf{x}_4), $$
each obeying (\ref{YSC})
$$ L_1(u_1^+,u_1) L_2(u_2^+, u_2) L_0(u_0^+, u_0) \Phi_{1,2,0}
= E \ \Phi_{1,2,0}, $$
 $$ L_0(v_0^+,v_0) L_3(u_3^+, u_3) L_4(u_4^+, u_4) \bar \Phi_{0,3,4}
= \bar E \ \bar \Phi_{0,3,4}. $$

We calculate the action of the order 4 monodromy on $\Phi_{1,2,3,4} $
$$  L_1(u_1^+,u_1)) L_2(u_2^+, u_2) L_3(u_3^+, u_3) L_4(u_4^+, u_4)
\Phi_{1,2,3,4} =  $$ $$
L_1(u_1^+,u_1)) L_2(u_2^+, u_2)
\int d \mathbf{x}_0
 \Phi_{1,2,0} \ \bar E (L_0(v_0^+,v_0) )^{-1} 
\bar \Phi_{0,3,4} . $$
Integration by parts moves $L_0^{-1}$ acting on $\bar \Phi$ to $L_0^{-1 T}$
acting on $\Phi$. 
The result is proportional to $\Phi_{1234}$ if
$$ (L_0(v_0^+,v_0) )^{-1 T } = const L_0 (u^+_0, u_0). $$

We evaluate the condition
$$
(L_0(v_0^+, v_0))^{-1} = const (L_0(u_0^+, u_0))^{T}, 
 $$
using (\ref{Lprojinv}) and (\ref{Lprojt})
$$ L_0(-u_0^+-1-n, -u_0-1) = const L_0( -v_0-1, -v_0^+-1), $$
and obtain the wanted conditions on the representation parameters
\be \label{conv}  u_0 = v_0^+, \ \ \ u_0^+ = v_0-n . \ee

\subsubsection{ Cyclicty}

If $\Phi$ obeys the YSC condition (\ref{YSC})  with $T(\mathbf{u})$  then it 
obeys this condition also with $T(\mathbf{u\p})$ with the change of the
sequence of points from $1, 2, ...,N$ to $2, ...,N,1$ and of the 
parameters by the cyclicity rule 
\be \label{cycl} \mathbf{u\p} =
\begin{pmatrix}
u_2 & ...& u_N & u_1 -n\\
u_2^+ & ...& u^+_N & u_1^+ -n 
\end{pmatrix}.
\ee

The cyclicity operation does not affect the weights associated with the points $I$.
We formulate and proof the cyclicity  rule (\ref{cycl})
in the situation before the weight projection (\ref{projL}) is done.

$$ L_1(u_1) L_2(u_2) ...L_N(u_N) \Phi = E \Phi $$
implies by inversion
$$ L_2(u_2) ...L_N(u_N) \Phi = E L_1^{-1}(u_1) \Phi $$
and by matrix transposition (\ref{Lt-1})
$$ L^t_N(u_N)...L^t_2(u_2)   = E ( L_1^{-1}(u_1))^t \Phi =
E ( L_1^{t}(u_1-n))^{-1} \Phi . $$
Multiplying by $  L_1^{t}(u_1-n) $ and applying the matrix transposition
once more we obtain
$$L_2(u_2) ...L_N(u_N)  L_1(u_1-n) \Phi = E \Phi . $$
In this way we have proven (\ref{cycl}).

The inversion (\ref{Lprojinv}) results in YSC with the reversed order of points.

\subsubsection{ The R operator method}

If $\Phi$ obeys the YSC condition (\ref{YSC})  with $T(\mathbf{u})$  then
we shall find operators 
$ R_{I, I+1}(u_I-u_{I+1}) $ and $ R_{I+1, I}(u^+_I-u^+_{I+1}) $ such that
their action on $\Phi$ leads to a YSC with monodromies of parameter sets
obtained from $\mathbf{u}$ by permuting adjacent entries,
$u_I$ and $u_{I+1}$ of the upper row in the first case and 
$u^+_I$ and $u^+_{I+1}$ of the lower row in the second.
By such $R$ operations we can construct more connected YSC.
The  weights associated with the points $I, I+1$
change 
from $2\ell_I = u_I^+ - u_I, 2\ell_{I+1} = u_{I+1}^+ - u_{I+1}$
by the action of $R_{I, I+1} $ to 
$2\ell_I\p = u_I^+ - u_{I+1}, 2\ell_{I+1} \p = u_{I+1}^+ - u_{I}$
or by the action of $R_{ I+1, I} $ to
$2\ell_I\pp = u_{I+1}^+ - u_{I}, 2\ell_{I+1}\pp = u_{I}^+ - u_{I+1}$.

\subsection{The Yang-Baxter operator}

In the fundamental Yang-Baxter relation (\ref{fundYB}) the $R$ matrix intertwines
the tensor product of fundamental $n$ dimensional representations.
We shall obtain a more general $R$ operator intertwining the tensor product
of JS type representations. It should obey the Yang-Baxter relation of the
form
\be \label{RLL} R_{12}(u-v) L_1(u) L_2(v) =  L_1(v) L_2(u)  R_{12}(u-v), \ee
$$ (L_1(u))_{ab} = \delta_{a b} u + \dd^1_a x_b^1,  \ \ \ 
 (L_2(u))_{ab} = \delta_{a b} u + \dd^2_a x_b^2. $$
The relation involves the matrix product $ L_1(u) L_2(v)$. 
$R_{12}$ is not a matrix, i.e. does not act on the fundamental representation
indices, but acts rather
on the two sets of canonical pairs $ x_a^1,\dd^1_a $ and $ x_a^2,\dd^2_a $.

We read (\ref{RLL}) as the defining relation of $R_{12}$ and solve it 
using the ansatz
$$ R_{12}(u) = \int \mathrm{d}c\, \phi(c)  
e^{-c (\underline{x}_1  \underline{\dd}_2)}. 
$$
We calculate the action of the shift operator on the product of $L$
matrices:
$$ e^{-c (\underline{x}_1  \underline{\dd}_2)}  \underline{\dd}_1 = 
(\underline{\dd}_1 + c \underline{\dd}_2 ) 
e^{-c (\underline{x}_1  \underline{\dd}_2)},
\qquad e^{-c (\underline{x}_1  \underline{\dd}_2)} \mathbf{x}_2 =
(\mathbf{x}_2 - c \mathbf{x}_1) \ e^{-c (\underline{x}_1  \underline{\dd}_2)},
$$
$$  e^{-c (\underline{x}_1  \underline{\dd}_2) } (L_1^+ (u) L_2^+
(v))_{ab}   e^{c (\underline{x}_1  \underline{\dd}_2)}= 
\big(L_{1 ac}^+ (u) + c \dd^2_{ad} x^1_{dc}  \big)
\big(L_{2 cb}^+ (v) - c \dd^2_{cd} x^1_{db} \big)
 =
$$ $$
\big( L_1^+ (v) L_2^+ (u) \big)_{ab} 
+ \big( L_2^+ (0) - L_1^+ (0)  \big)_{ab} \{ [u-v + c
(\underline{x}_1 \cdot \underline{\dd}_2) ] 
- c [ u-v -1 + c (\underline{x}_1  \underline{\dd}_2) ] \}. 
$$
The condition that the contribution of the second term 
vanishes implies a differential equation
on $\phi(c) $, because
$$ \int \mathrm{d}c\, \phi(c)  \cdot c\cdot (\underline{x}_1  \underline{\dd}_2)
e^{-c (\underline{x}_1  \underline{\dd}_2)} = 
- \int \mathrm{d}c\, \phi(c) \cdot  c \cdot \dd_c\,
e^{-c (\underline{x}_1  \underline{\dd}_2)} = 
\int \mathrm{d}c\, \dd_c  (c \phi(c) )  \, 
e^{-c (\underline{x}_1  \underline{\dd}_2)}. $$ 
In the last step we have assumed that the integration by parts is done
without boundary terms as it holds for closed contours.
Thus the condition on $\phi$ is
$$ \dd_c ( c \phi(c)) +  (u-v) \phi(c) = 0, $$
and it is solved by
$$ \phi(c) = \frac{1}{c^{1 + u-v} }. $$
The condition of vanishing of the third term can be written as
$$ 0 = \dd_c( c^2 \phi(c)) + c (u-v-1)\phi(c) = 
c [ \dd_c( c \phi(c)) + (u-v)\phi(c) ].
$$
We see that it does not imply a further condition on $\phi(c)$.
Thus we have proved that 
\be \label{R12} 
R_{12}(u) = \int \frac{\mathrm{d}c}{c^{1+u} } 
e^{-c (\underline{x}_1  \underline{\dd}_2 )}
\ee
 obeys the Yang-Baxter relation provided the simple rule of integration by parts.
If the latter rule is different then the indicated procedure leads to the
appropriate modification. 
 
Now we derive two Yang-Baxter relations for the projected $L$ operators
(\ref{projL}). We notice that the shift operation included in
the above solution of $R_{12}$ affects the dilatation operators
$$
 e^{-c(\underline{x}_1  \underline{\dd}_2) } (\underline{x}_1
\underline{\dd}_1) e^{c(\underline{x}_1  \underline{\dd}_2)
 } = (\underline{x}_1  \underline{\dd}_1) + c(\underline{x}_1
\underline{\dd}_2), \ \ 
 e^{-c(\underline{x}_1  \underline{\dd}_2) } (\underline{x}_2  \underline{\dd}_2) 
e^{c(\underline{x}_1  \underline{\dd}_2) } = 
(\underline{x}_2  \underline{\dd}_2) - c(\underline{x}_1
\underline{\dd}_2).
$$ 
 
By  partial integration we obtain 
$$ \int dc \phi(c)  c (\underline{x}_1  \underline{\dd}_2)  
e^{-c(\underline{x}_1  \underline{\dd}_2)} =
- \int dc \phi(c)  c \dd_c  e^{-c(\underline{x}_1  \underline{\dd}_2)} = 
$$ $$
\int dc (\dd_c c \varphi(c) )  e^{-c(\underline{x}_1  \underline{\dd}_2)} = 
(u-v) \int dc \varphi(c)    e^{-c(\underline{x}_1  \underline{\dd}_2)} . $$
In this way we have 
$$ R_{12}(u-v) \left ( v + (\underline{x}_1  \underline{\dd}_1) \right )
R_{12}^{-1}(u-v) 
= u+ (\underline{x}_1  \underline{\dd}_1), \ \ 
R_{12}(u-v) \left ( u + (\underline{x}_2  \underline{\dd}_2) \right )
R_{12}^{-1}(u-v) 
= v+ (\underline{x}_2  \underline{\dd}_2)
$$
or 
\be \label{RPi} R_{12}(u-v)\Pi_1(2\ell_1) = \Pi_1(2\ell_1+ u-v)
R_{12}(u-v),\ \ \
R_{12}(u-v)\Pi_2(2\ell_2) = \Pi_2(2\ell_2- u+v) R_{12}(u-v). \ee

This implies that  (\ref{RLL}) results for 
the projected $L$ in
a similar relation with the permutation of $u$ and $v$ but coincidence
of the additional arguments  on both sides.
\be \label{RLLproj1}  R_{12}(u-v) L_1(u^+, u) L_2(v^+, v) =  L_1(u^+, v) L_2(v^+, u)
 R_{12}(u-v). \ee
 Now we apply the inversion relation of the projected $L$ (\ref{Lprojinv}) to obtain 
the Yang-Baxter relation with permutation of the first arguments.
We have from the last relation (\ref{RLLproj1})
$$  R_{12}(u-v) (L_2(v^+,v))^{-1} (L_1(u^+,u))^{-1} = 
(L_2(v^+,u))^{-1}  (L_1(u^+,v))^{-1}  R_{12}(u-v). $$
The inversion relation (\ref{Lprojinv}) implies
$$  R_{12}(u-v) L_2(-v-1, -v^+-1) L_1(-u-1, -u^+-1) = $$ $$
L_2(-u-1, -v^+-1)  L_1(-v-1 ,-u^+-1)  R_{12}(u-v). $$
We interchange the labels $1 \leftrightarrow 2$,
$$  R_{21}(u-v) L_1(-v-1, -v^+-1) L_2(-u-1, -u^+-1) = $$ $$
L_1(-u-1, -v^+-1)  L_2(-v-1 ,-u^+-1)  R_{21}(u-v), $$
and substitute the representation parameters as
$$ u\p = -v^+-1, v\p = -u^+-1, {u\p}^+= -v-1, {v\p}^+ = -u-1 $$
to obtain 
\be \label{RLLproj2}
  R_{21}({u\p}^+ -{v\p}^+) L_1({u\p}^+, u\p) L_2({v\p}^+, v\p) 
=  L_1({u\p}^+, v\p) L_2({v\p}^+, u\p)  R_{21}({u\p}^+ -{v\p}^+). \ee
Now the primes can be omitted.
We have obtained the Yang-Baxter relations for the $L$ operators projected
to definite weights in the convenient form characterized by the
permutation of the first or the second arguments. This results in the 
$R$ operation rule for the YSC above.

The repeated application of this rule allows to construct
YSC starting from the trivial products of delta distributions 
in $K$ of the $N$ points and constant functions in the other points,
e.g.
$$ \Phi_{K,N; 0} (\mathbf{x}_1, ..., \mathbf{x}_N) =
\Pi_{L=1}^K \delta ((\mathbf{x}_L). $$ 
The expression of the $R$ operators in terms of 
a contour integral over a shift operator (\ref{R12}) leads to the general form
\be \label{link} \Phi_{K,N} (\mathbf{x}_1, ..., \mathbf{x}_N) =
\int \varphi(c) \Pi_{L=1}^K dc_{L, K+1}...dc_{L,N}
\delta (\mathbf{x}_L- c_{L, K+1} \mathbf{x}_{K+1} - c_{L, N} \mathbf{x}_N).
\ee
The integration variables can be regarded as coordinates on the
Gra\ss mann variety $\mathcal{M}_{K,N} $. The integrand function
$\varphi(c)$ contains the information about the particular YSC.
The starting YSC $\Phi_{K,N; 0}$, where the points are disconnected,
 can be written in this form too
with $\varphi(c)$ being a product of delta distributions in the
$c_{L, M} $ .
If the points are maximally connected, then the rectangular matrix
with the elements $c_{L, M}, L=1, ...K; M= 1, ..., N$ has maximal rank,
the integration extends over the maximal Schubert cell of 
$ \mathcal{M}_{K,N}  $.

\subsection{The helicity representation}

We have noticed that the two forms $L^+, L^-$ of the JS presentation of the
$L$ operators (\ref{L+-}) are related by the  elementary canonical transformation
(\ref{canon}). 
There the transformation acts uniformly for the index values $a = 1, ...n$. 
Now we restrict ourselves to
the case of even $n= 2m$. We split the index range into the parts
$1, ..., m$ and $m+1, ..., 2m$ and label the first by $\alpha = 1, .., m$
and the second by $\alphadot = 1, ..., m$.
 With a transformation modified
in this respect, i.e. non-trivial for the index value $a = \alpha = 1, ...,
m$ only, 
\be \label{canon1}
 \begin{pmatrix}
x_{I,\alpha} \\ \dd_{I,\alpha} 
\end{pmatrix} \rightarrow
 \begin{pmatrix}
\dd^{\lambda}_{I, \alpha} \\ -\lambda_{I, \alpha} 
\end{pmatrix},  \ \ \ 
\begin{pmatrix}
x_{I,m+\alphadot} \\ \dd_{I,m+\alphadot} 
\end{pmatrix} \rightarrow
\begin{pmatrix}
\bar \lambda_{I, \alphadot} \\ \bar \dd^{\lambda}_{I, \alphadot} 
\end{pmatrix}, 
\ee
applied to $L^+_I$ (\ref{L+-}) we are lead to the helicity form  
$L^{\lambda}_I$ 
appearing in $m \times m$ blocks (for any $I$ )
$$ L^{\lambda}_{ \alpha, \beta} = - \lambda_{\alpha}
\dd^{\lambda}_{\beta}, \ \ \  
  L^{\lambda}_{ \alpha, \betadot} 
= - \lambda_{  \alpha}\bar \lambda_{ \betadot},
$$ $$
 L^{\lambda}_{ \alphadot , \beta} = \bar
\dd^{\lambda}_{\alphadot}\dd^{\lambda}_{ \beta},
\ \ \ 
L^{\lambda}_{ \alphadot , \betadot} 
=\bar \dd^{\lambda}_{ \alphadot} \bar \lambda_{ \betadot}. 
$$

The transformation applied to the dilatation operator 
$(\underline{x} \underline{\dd} )$ results in
$$ (\underline{x} \underline{\dd} ) \to 
- \dd^{\lambda}_{ \alpha} \lambda^{ \alpha} + 
\bar \lambda_{\alphadot} \bar \dd^{\lambda \  \alphadot} 
= -m +2 \hat h
$$ 
We notice that (\ref{canon1}) is for $m=2$ the inversion of 
(\ref{cantrans1}), (\ref{cantrans2}). 
In view of the application to scattering we call 
$\hat h$ helicity. 
The relation of its eigenvalues to the dilatation weight is
\be \label{2h2l}  2h = m + 2\ell . \ee 

The expression for $R^{\lambda}_{12}$ is obtained from the one for
$R_{12}(u)$ (\ref{R12}) by substituting according to the canonical
transformation (\ref{canon1})
\be \label{x1d2}
 (\underline{x}_1  \underline{\dd}_2) \to 
-\dd^{\lambda}_{1 \alpha} \lambda_{2}^{\alpha} 
+ \bar \lambda_{1 \alphadot} \bar \dd^{\lambda \alphadot}_{2}.
\ee

The YSC condition (\ref{YSC}) implies in particular
\be \label{momentum} \sum_I L^{\lambda}_{I \alpha \alphadot} \ \Phi^{\lambda} = 
\sum_I \lambda_{I \alpha} \bar \lambda_{\alphadot} \ \Phi^{\lambda}
= 0 . \ee
The transformation of $\Phi_{K, N} $ (\ref{link}) to the helicity representation
results in
\be \label{linkhel} 
\Phi^{\lambda}_{K,N} (\lambda_1, \bar \lambda_1 ..., \lambda_N, \bar \lambda_N) =
\int \varphi(c) \Pi_{L=1}^K dc_{L, K+1}...dc_{L,N} $$ $$
\delta^{(m)} (\bar \lambda_L- c_{L, K+1} \bar \lambda_{K+1} - c_{L, N} \bar
\lambda_N) 
\Pi_{M=K+1}^N \delta^{(m)} ( \lambda_M + c_{1,M} \lambda_1 + ...+c_{K, M}
\lambda_K ).
\ee
Instead of $K$ factors of $n=2m$ dimensional delta distributions we have now
$N$ factors of $m$ dimensional ones. The integrand function $ \varphi(c)$
is not changed.
(\ref{momentum}) follows from the linear equations implied by the
arguments of the delta distributions. This means,
the factor $ \delta^{(2m)} (\sum_I \lambda_{I \alpha} \bar
\lambda_{I \alphadot})$ can be separated.

\subsection{Examples of YSC constructions}

\subsubsection{Two and three points}

By one $R$ operation we obtain the 2-point YSC
$$
\Phi^{-+}_{12}(\mathbf{x}_1, \mathbf{x}_2) = R_{21}(u_1^+ -u_2) \delta(\mathbf{x}_1)
= \int \frac{dc_{12} }{c_{12}^{1+ u_1^+ - u_2} } \delta(\mathbf{x}_1
- c_{12} \mathbf{x}_2). $$ 
Let us abbreviate in $\Phi$ the dependence on the points $ \mathbf{x}_I$
by subscripts $I$.

We construct 3-point correlators starting from the trivial ones
$$ \Phi_{123, 0}^{--+} = \delta(\mathbf{x}_1) \delta(\mathbf{x}_2), \ \ \ 
\Phi_{123, 0}^{-++} = \delta(\mathbf{x}_1). $$
The corresponding monodromy is characterized by the set of parameters
$\mathbf{u}$ 
conveniently displayed by rows of $u_I$ and $u^+_I$ (\ref{projmono}). We abbreviate here 
parameters by their position indices $I$ eventually supplemented by a superscript $+$
for indicating the shift by $-n$.
In case $\Phi_{123, 0}^{--+}$ we have
$$ \mathbf{u}^{--+ (0)} =
\begin{pmatrix}
1&2&3&\\
1^+&2^+&3 
\end{pmatrix},
$$
and in the case $\Phi_{123, 0}^{-++}$
$$ \mathbf{u}^{-++ (0)} =
\begin{pmatrix}
1&2&3&\\
1^+&2&3 
\end{pmatrix}.
$$
By one $R$ operation we obtain
$$  \Phi_{123, 1}^{--+} = R_{31}(u_1^+-u_3)  \Phi_{123, 0}^{--+} =
\int \frac{c_{13} }{ c_{13}^{1+ u^+_1-u_3} } \delta(\mathbf{x}_1
- c_{13}\mathbf{x}_3 )\delta(\mathbf{x}_2)$$
with 
$$ \mathbf{u}^{--+ (1)} =
\begin{pmatrix}
1&2&3&\\
3&2^+&1^+  
\end{pmatrix}.
$$
 By a second $R$ operation we obtain the completely connected correlator.
\be \label{Phi3--+} 
 \Phi_{123, 2}^{--+} = R_{32}u^+_2-u^+_1)R_{31}(u_1^+-u_3)  \Phi_{123, 0}^{--+} = 
\int \frac{dc_{13} }{ c_{13}^{1+ u^+_1-u_3} } \frac{dc_{23} }{ c_{23}^{1+
u^+_2-u^+_1} }
\delta(\mathbf{x}_1
- c_{13}\mathbf{x}_3 )\delta(\mathbf{x}_2 - c_{23}\mathbf{x}_3 ) \ee
with 
\be \label{u3--+} \mathbf{u}^{--+ (2)} =
\begin{pmatrix}
1&2&3&\\
3&1^+&2^+  
\end{pmatrix}.
\ee

For the other case we have
$$ \Phi_{123, 1}^{-++} = R_{31} (u_1^+-u_3) \delta(\mathbf{x}_1) = 
\int \frac{dc_{13}}{c_{13}^{1+ u_1^+ - u_3} } \delta(\mathbf{x}_1 - c_{13} \mathbf{x}_3 ) 
$$
with 
$$  \mathbf{u}^{-++ (1)} =
\begin{pmatrix}
1&2&3&\\
3&2&1^+  
\end{pmatrix}.
$$
The completely connected correlator is obtained in the next step as
\be \label{Phi3-++}
 \Phi_{123, 2}^{-++} = R_{21}(u_3-u_2) R_{31} (u_1^+-u_3) \delta(\mathbf{x}_1) =
\int \frac{dc_{13}}{c_{13}^{1+ u_1^+ - u_3} } \frac{dc_{12}}{c_{12}^{1+ u_3 -
u_2} }
\delta(\mathbf{x}_1 - c_{12} \mathbf{x}_2 - c_{13} \mathbf{x}_3) 
\ee
with
\be \label{u3-++} \mathbf{u}^{-++ (2)} =
\begin{pmatrix}
1&2&3&\\
2&3&1^+  
\end{pmatrix}.
\ee

\subsubsection{Four points}

We illustrate the construction of a connected 4-point YSC 
in two ways, first by R operations
on the product of 2-point YSC and second by the convolution
of two 3-point YSC followed by a R operation. 

We start from the product of 2-point YSC,
$$ \Phi_{1234, 2} = \Phi_{12} \Phi_{34} = 
R_{21}(u_1^+ -u_2) R_{43}(u_3^+ - u_4) \delta(\mathbf{x}_1)
\delta(\mathbf{x}_3) = $$ $$
\int \frac{dc_{12} dc_{34}}{c_{12}^{1+u_1^+-u_2} c_{34}^{1+u_3^+-u_4} }
\delta(\mathbf{x}_1 -c_{12}\mathbf{x}_2)
\delta(\mathbf{x}_3-c_{34}\mathbf{x}_4). 
$$
The corresponding monodromy is characterised by the set of parameters
\be \label{u2}  \mathbf{u}^{(2)} =
\begin{pmatrix} 
1&2&3&4 \\
2 & 1^+ & 4 & 3^+
\end{pmatrix}.
\ee
To obtain a completely connected 4-point YSC we act by $R_{23} R_{32} $.
The appropriate arguments are read off from (\ref{u2}).
The action by $ R_{32} $ results in
$$ \Phi_{1234, 3} = R_{32}(u_1^+-u_4)  \Phi_{1234, 2} = $$ $$
\int \frac{ dc_{23} dc_{12} dc_{34}}{
 c_{23}^{1+u_1^+-u_4} c_{12}^{1+u_1^+-u_2} c_{34}^{1+u_3^+-u_4} }
\delta(\mathbf{x}_1 -c_{12}\mathbf{x}_2- c_{12}c_{23} c_{34} \mathbf{x}_4) 
\delta(\mathbf{x}_3- c_{34}\mathbf{x}_4). 
$$
The substitution $c_{14} = c_{12}c_{23} c_{34}$ leads to
$$ \Phi_{1234, 3} =
\int \frac{ dc_{23} dc_{12} dc_{34}}{
 c_{14}^{1+u_1^+-u_4} c_{12}^{1+u_4-u_2} c_{34}^{1+u_3^+-u_1^+} }
\delta(\mathbf{x}_1 -c_{12}\mathbf{x}_2- c_{14} \mathbf{x}_4) 
\delta(\mathbf{x}_3 - c_{34}\mathbf{x}_4). 
$$
The monodromy parameters are
\be \label{u3}  \mathbf{u}^{(3)} =
\begin{pmatrix} 
1&2&3&4 \\
2 & 4 & 1^+ & 3^+
\end{pmatrix}.
\ee
This correlator is still not completely connected.
The action by $R_{23}$ leads to the completely connected correlator
$$ \Phi_{1234, 4} = \int \frac{dc_{32} dc_{23} dc_{12} dc_{34}}{
c_{32}^{1+u_2-u_3} c_{14}^{1+u_1^+-u_4} c_{12}^{1+u_4-u_2}
c_{34}^{1+u_3^+-u_1^+} }
\delta(\mathbf{x}_1 -c_{12}\mathbf{x}_2- c_{14} \mathbf{x}_4) 
\delta(\mathbf{x}_3-c_{32}\mathbf{x}_2 - c_{34}\mathbf{x}_4), 
$$
with
\be \label{u4}  \mathbf{u}^{(4)} =
\begin{pmatrix} 
1&3&2&4 \\
2 & 4 & 1^+ & 3^+
\end{pmatrix}.
\ee

Now we show how the same result is obtained by the convolution
method. 

We consider the  convolution of 3-point YSC (\ref{Phi3-++}) and
(\ref{Phi3--+}),
$$ \int d\mathbf{x}_0 \Phi_{120,2}^{-++}(\mathbf{u})  
\Phi_{034,2}^{--+}(\mathbf{v}). 
$$  
The condition on the spectral parameters is obtained from (\ref{conv})
taking into account the parameter sets (\ref{u3-++}) and (\ref{u3--+}),
\be \label{v|u}
v_0-n = u_1-n, u_0-n = v_4-n. \ee
Thus we have
$$
\int \frac{dc_{12} dc_{10} dc_{04} dc_{34} }{ c_{12}^{1+v_4-u_2} 
c_{10}^{1+u_1^+-v_4} c_{04}^{1+u_1-n -v_4} c_{34}^{1+u_2^+-u_1^+} }
\delta( \mathbf{x}_1 - \tilde c_{12} \mathbf{x}_2 - \tilde c_{10}c_{04}
\mathbf{x}_4) \delta (\mathbf{x}_3 - c_{34} \mathbf{x}_4).
 $$ 
The substitution $c_{14} = c_{10} c_{04}$ leads to
$$
\int \frac{dc_{04}}{c_{04}} \int \frac{dc_{12} dc_{14} dc_{34} }{
c_{12}^{1+v_4-u_2}c_{14}^{1+u_1^+-v_4} c_{34}^{1+u_2^+-u_1^+} }
\delta( \mathbf{x}_1 - \tilde c_{12} \mathbf{x}_2 - \tilde c_{14}
\mathbf{x}_4) \delta (\mathbf{x}_3 - c_{34} \mathbf{x}_4).
 $$ 
The first integral can be regarded as an irrelevant factor.
The result can be represented as generated directly by three
$R$ operators as in the next-to-last step of $R$ operations above
$$ R_{32}(u_1^+-u_4) R_{21}(u_1^+ -u_2) R_{43}(u_3^+ - u_4) \delta(\mathbf{x}_1)
\delta(\mathbf{x}_3) 
$$ 
with the substitution $c_{14}= c_{12}c_{23} c_{34}$, i.e.
it coincides with $\Phi_{1234, 3}$ and the parameter set with (\ref{u3}).
As above the 
 completely connected correlator (\ref{u4}) is obtained  by the action
of $R_{23}$. Thus the reconstruction of this 4-point YSC by convolution of 
two 3-point YSC can be summarized as
\be \label{fusion33}
\Phi_{1234, 4} = R_{23}(u_2-u_3) \int   
d\mathbf{x}_0 \Phi_{120,2}^{-++}(\mathbf{u})  
\Phi_{034,2}^{--+}(\mathbf{v})|_{v|u},
\ee
where $v|u$ advises the substitution (\ref{v|u}).

The points of the correlators seem to be of two types marked
 by signature $\pm$. Those $\mathbf{x}_I$
entering the initial trivial  correlator $\Phi_{..., 0}$ by $\delta(\mathbf{x}_I)$
we have marked by $-$ and the others by $+$. In particular, 
our examples above of 4-point YSC
carry the signature $-+-+$. In completely connected YSC this signature
assignment is inessential. This holds in general, we show this in the case of 4 points. 

 The standard $-+-+$ form 
$$ 
\Phi_{1234}^{-+-+} = \int dc \varphi^{-+-+}(c_{12}, c_{14},c_{32},c_{34}) 
\delta(\mathbf{x}_1 -c_{12}\mathbf{x}_2- c_{14} \mathbf{x}_4) 
\delta(\mathbf{x}_3-c_{32}\mathbf{x}_2 - c_{34}\mathbf{x}_4) 
$$
can be rewritten as e.g.
$$ 
\Phi_{1234}^{--++} = \int d c\p \varphi^{--++}( c\p) 
\delta(\mathbf{x}_1 - c\p_{13}\mathbf{x}_3-  c\p_{14} \mathbf{x}_4) 
\delta(\mathbf{x}_2- c\p_{23}\mathbf{x}_3 -  c\p_{24}\mathbf{x}_4) 
$$ 
where 
\be \label{phitophi}
 \varphi^{--++}(c\p_{13}, c\p_{14}, c\p_{23}, c\p_{24}) = c_{23}^{\prime
n-4} 
\varphi^{-+-+}(\frac{c\p_{13}}{c\p_{23} }, \frac{c\p_{14} c\p_{23} - c\p_{24}
c\p_{13}}{ c\p_{23}}, \frac{1}{c\p_{23} } , - \frac{c\p_{24}}{c\p_{23}} ).
\ee

Therefore the  maximally connected correlator (\ref{u4})
transforms to the $--++$ form as
\be \label{Phi4--++} 
\Phi_{1234, 4}^{--++} = \int \frac{d c\p_{13} dc\p_{14} dc\p_{23} dc\p_{24}
 }{ (c\p_{13})^{1+u_4-u_2} (c\p_{24})^{1+u_3^+ - u_1^+} (c\p_{14} c\p_{23} - c\p_{24}
 c\p_{13})^{1+u_1^+ -u_4} }
\ee $$\delta(\mathbf{x}_1 - c\p_{13}\mathbf{x}_3-  c\p_{14} \mathbf{x}_4) \ 
\delta(\mathbf{x}_2- c\p_{23}\mathbf{x}_3 -  c\p_{24}\mathbf{x}_4) 
$$ 
Both expressions $\Phi_{1234, 4}^{-+-+}$ and $\Phi_{1234, 4}^{--++}$
are equivalent, in particular they obey the YSC condition (\ref{YSC})
with the same monodromy of the parameters (\ref{u4}).
The latter expression (\ref{Phi4--++}) can be obtained directly from 
$\Phi_{1234, 0}^{--++} = \delta(\mathbf{x}_1) \delta(\mathbf{x}_2) $
by $R$ operations 
\be \label{Phi4--++R} 
\Phi_{1234, 4}^{--++} = R_{12}(v_1-v_2) R_{21}(v_4-v_3) R_{32}(v_2^+ -v_3)
R_{41}(v_1^+ -v_4) \  \delta(\mathbf{x}_1) \delta(\mathbf{x}_2) 
\ee
and corresponds to the monodromy with the parameters $\mathbf{v}$
abbreviated by the pattern
\be \label{v4} \mathbf{v}^{(4)} =
\begin{pmatrix}
2&1&3&4 \\
3&4&2^+&1^+
\end{pmatrix}.
\ee
The parameters $\mathbf{u}^{(4)}$ and $\mathbf{v}^{(4)}$ are related as
$ u_1 = v_2, u_2 = v_3, u_3 = v_1, u_4 = v_4 $. 
Indeed, r.h.s of (\ref{Phi4--++R}) reads explicitly
$$ \int \frac{dc_{21} dc_{12} dc_{23} dc_{14} }{c_{21}^{1+v_1-v_2} 
c_{12}^{1+v_4-v_3} c_{23}^{1+v_1^+-v_3} c_{14}^{1+ v_1^+ -v_4} }
$$ $$
\delta(\mathbf{x}_1 -c_{12} c_{23} \mathbf{x}_3 - c_{14} \mathbf{x}_4)
\delta (\mathbf{x}_2 - (1+c_{12} c_{21} +1) c_{23} \mathbf{x}_3 -
c_{21}c_{14} \mathbf{x}_4)
$$
The transformation to the standard $--++$ form (\ref{phitophi}
results in
\be \label{Phi4--++v} 
\Phi_{1234, 4}^{--++} = \int \frac{dc_{13}\p dc_{14}\p dc_{23}\p dc_{24}\p }{
c_{13}^{\prime 1+v_4-v_3} c_{24}^{\prime 1+ v_1-v_2} 
 (c\p_{14} c\p_{23} - c\p_{24} c\p_{13})^{1+v_2^+ -v_4} }
\ee
$$\delta(\mathbf{x}_1 - c\p_{13}\mathbf{x}_3-  c\p_{14} \mathbf{x}_4) \ 
\delta(\mathbf{x}_2- c\p_{23}\mathbf{x}_3 -  c\p_{24}\mathbf{x}_4) 
$$ 
with the monodromy parameters (\ref{v4}). 
We shall use $\Phi_{1234; 4}^{--++}$ (\ref{Phi4--++R}) and the intermediate YSC of this
 R operation construction $\Phi_{1234; 2}^{--++} $ and $\Phi_{1234; 3}^{--++}$.

\subsubsection{Crossing}

We recall that
the parameters $u^+_I, u_I$ are related to the dilatation weight of the YSC
at the point $I$ as $2\ell_I = u_I^+ -u_I$. The set of weights in 
a correlator $\Phi_{K,N} $ obeys the constraint
\be \label{sumlI} \sum_{I=1}^N 2\ell_I + n K  = 0. \ee
The weights of the 4-point YSC $\Phi_{1234, 4}$ constructed above
are more restricted,
\be \label{l13l24} 
2\ell_1 + 2\ell_3 + n = 0, \ \ 2\ell_2 + 2 \ell_4 + n = 0. \ee
We intend to construct a 4-point YSC where such constraints connect
the points $1 , 4$ and $2, 3$ instead.

For this we apply additional $R$ operators to the YSC (\ref{v4}), (\ref{Phi4--++v}),
$$ R_{43} (v_2-v_1) R_{34} (v_3-v_4) \Phi_{1234;4}^{--++4}. $$
The action affects the arguments of the delta distributions
(\ref{Phi4--++v}),
$$ \delta ( \mathbf{x}_1 - (c_{13} -c_{14} \bar c_{42}) \mathbf{x}_3
-( c_{14}(1+\bar c_{43} \bar c_{34} ) - c_{13} \bar c_{34}) \mathbf{x}_4)
\ 
 \delta ( \mathbf{x}_2 - (c_{23} -c_{24} \bar c_{42}) \mathbf{x}_3
-( c_{24}(1+\bar c_{43} \bar c_{34} ) - c_{23} \bar c_{34}) \mathbf{x}_4).
$$
The standard form (\ref{link}) is recovered by the transformation of the
 variables $ c_{I 3}, c_{I 4} $ to $\tilde c_{I 3}, \tilde c_{I 4} $
($I=1,2$ ),
$$
\begin{pmatrix}
\tilde c_{I3} \\
\tilde c_{I4}\\ 
\end{pmatrix} 
= 
\begin{pmatrix}
1 & - \bar c_{43} \\
- \bar c_{34} & 1+ \bar c_{43} \bar c_{34} \\
\end{pmatrix} 
\begin{pmatrix}
 c_{I3} \\
 c_{I4}\\ 
\end{pmatrix}, 
$$
by the matrix depending on the  variables $ \bar c_{34}, \bar c_{43}$,
corresponding to the additional $  R_{43}, R_{34} $ respectively. 
 
We transform the correlator integrand $\varphi(c)$ of $\Phi_{1234,4}^{--++}$ 
(\ref{Phi4--++v}) to the new variables
$$ (\varphi (c)^{-1}  \cdot \bar c_{34}^{1+v_2-v_1} \bar c_{43}^{1+
v_3-v_4} = 
( (1+\bar c_{43} \bar c_{34} ) \tilde c_{13} \bar c_{43} \tilde c_{14})^{1+
v_4-v_3} \ ( \bar c_{34} \tilde c_{23} + \tilde c_{24} )^{ 1+v_1-v_2} $$ $$
  ( \tilde c_{13} \tilde c_{24} -  \tilde c_{23} \tilde c_{14} ) ^{1+ v_2^+
 - v_4}  
\cdot \bar c_{34}^{1+v_2-v_1} \bar c_{43}^{1+
v_3-v_4} = $$ $$ 
( \tilde c_{13} \tilde c_{24} -  \tilde c_{23} \tilde c_{14} ) ^{1+ v_2^+
 - v_4}   
\bar c_{34}^2 ( \tilde c_{23} + \frac{\tilde c_{24}}{\bar c_{34} } )^{1+
   v_1-v_2} \ \ 
\bar c_{43}^2 ( \tilde c_{14} + \frac{\tilde c_{13}}{\bar c_{43} }
(1+\bar c_{43} \bar c_{34} ) )^{1+
   v_4-v_3}.
$$
We do the integrals over $ \bar c_{43}, \bar c_{34} $.
$$ \int \frac{d\bar c_{43} }{ \bar c_{43}^2 
( \tilde c_{14} + \frac{\tilde c_{13}}{\bar c_{43} }
(1+\bar c_{43} \bar c_{34} ) )^{1 +  v_4-v_3} } = 
\tilde c_{14}^{-1-v_4+ v_3}  \int \frac{dc\p_{43} }{(
1+ \frac{\tilde c_{13}}{\tilde c_{14}} \bar c_{34} + c\p_{34} 
\frac{\tilde c_{13}}{\tilde c_{14}} )^{1+ v_4-v_3} }.
$$
We have substituted $ c\p_{43} = \bar c_{43}^{-1} $ and in the next step
we substitute $ c_{43}^{\prime \prime} = c\p_{43} \frac{\tilde c_{13}}{\tilde c_{14}}
+1 + \bar c_{34} \frac{\tilde c_{13}}{\tilde c_{14}} $ and obtain
$$ \frac{1}{ \tilde c_{13} \tilde c_{14}^{v_4 - v_3} } 
\int \frac{c_{43}^{\prime \prime} }{ c_{43}^{\prime \prime 1+ v_4 - v_3} }.
 $$
Similarly, we substitute $ c_{34}\p = \frac{\tilde c_{24} }{ \tilde c_{23}
\bar c_{34} } $,
$$ \int \frac{d\bar c_{34} }{ \bar c_{34}^2 ( \tilde c_{23} + 
\frac{\tilde c_{24} }{ \bar c_{34}} )^{1+ v_1 - v_2} } = 
 \frac{1}{ \tilde c_{24} \tilde c_{23}^{v_1 - v_2} } 
\int \frac{c_{34}^{\prime } }{ c_{43}^{ \prime 1+ v_1 - v_2} }.
$$
The integrals over $c_{43}^{\prime \prime}, c_{34}\p $ result
in undefined but irrelevant factors. 
As the result the correlator 
$ R_{43} (v_2-v_1) R_{34} (v_3-v_4) \Phi^{--++,4} $
has the standard form (\ref{link})
with the integrand $\tilde \varphi $,
\be \label{Phi4c} (\tilde \varphi ( c) )^{-1} =
 c_{13} c_{14}^{v_4-v_3} c_{24} c_{23}^{ v_1-v_2} 
(c_{23}c_{14} - c_{13} c_{24} )^{1+ v_2^+ - v_4},  \ee
and the parameters of the $L$ operators in the monodromy $L_1 L_2 L_3 L_4$
are
abbreviated by  
\be \label{v4c} \mathbf{v}^{(4)c} =
\begin{pmatrix}
2&1&4&3 \\
3&4&1^+&2^+
\end{pmatrix}.
\ee
Now the weight restriction reads 
\be \label{l14l23} 2\ell_1 + 2\ell_4 +n =0, \ \ \ 
2\ell_2 + 2\ell_3 +n =0 . \ee 
If one would like to have the YSC with  the related weights to be adjacent
the cyclicity relation (\ref{cycl}) is to be applied. 
The same correlator obeys the YSC relation with the monodromy
$L_2 L_3 L_4 L_1$ where the parameter arguments are read off from 
\be \label{v4cc} \mathbf{v}^{(4)cc} =
\begin{pmatrix}
1&4&3&2-n \\
4&1^+&2^+&3-n
\end{pmatrix}.
\ee
Here $-n$ in the last column means that  the 
last factor with the explicit arguments is $L_1(v_3-n, v_2-n) $.

\subsection{Integral operators from symmetric correlators } 

We consider 
integral operators with YSC as kernels. Such operators 
act in a highly symmetric way, defining
homomorphisms not only of the $g\ell(n)$ Lie algebra action on the functions 
(global symmetry) but of the
related Yangian  algebra. Let us illustrate this in the case of
operators on functions of two points with 4-point YSC as kernels \cite{FK16}.

We define the operator $Q$ in action on $\psi(\mathbf{x}_1, \mathbf{x}_2) $
as
\be \label{defQ} Q\psi(\mathbf{x}_1, \mathbf{x}_2)  = 
\int d\mathbf{x}\p_1 d\mathbf{x}\p_2 \psi(\mathbf{x}\p_1, \mathbf{x}\p_2) 
\Phi(\mathbf{x}\p_2, \mathbf{x}\p_1,\mathbf{x}_2, \mathbf{x}_1 )
\ee
with the YSC condition (\ref{YSC}) ( relabelling $1 2 3 4 \to 2\p 1\p 2 1 $)
on the kernel
$$ L_{2\p}(u_2^{\prime +}, u_2\p)  L_{1\p}(u_1^{\prime +}, u_1\p) 
 L_2(u_2^{ +}, u_2)  L_1(u_1^{ +}, u_1) 
\Phi(\mathbf{x}\p_2, \mathbf{x}\p_1,\mathbf{x}_2, \mathbf{x}_1 )
= E(\mathbf{u})  \Phi(\mathbf{x}\p_2, \mathbf{x}\p_1,\mathbf{x}_2, \mathbf{x}_1 ).
$$
The involved monodromy has the parameter pattern
\be \label{uprime}
 \mathbf{u}\p =
\begin{pmatrix}
2\p & 1\p & 2& 1 \\
2^{\prime +}& 1^{\prime +} & 2^+ & 1^+ 
\end{pmatrix}.
\ee

We show that $Q$ obeys 
\be \label{QLL} L_2(u_2^{ +}, u_2)  L_1(u_1^{ +}, u_1) Q = 
F((\mathbf{u}\p) 
Q L_1(u_1\p -n, u_1^{\prime +})  L_2(u_2\p -n, u_2^{\prime +}),
\ee  $$
F(\mathbf{u}\p) = \frac{E(\mathbf{u}\p)}{u_1\p (u_1^{\prime +}+1) 
u_2\p (u_2^{\prime +}+ 1)}.
$$
Indeed by the YSC condition on the kernel, 
$$ L_2(u_2^{ +}, u_2)  L_1(u_1^{ +}, u_1) Q = E(\mathbf{u}) 
\int d\mathbf{x}\p_1 d\mathbf{x}\p_2 \psi(\mathbf{x}\p_1, \mathbf{x}\p_2) 
  L^{-1}_{1\p}(u_1^{\prime +}, u_1\p) L^{-1}_{2\p}(u_2^{\prime +}, u_2\p) 
\Phi(\mathbf{x}\p_2, \mathbf{x}\p_1,\mathbf{x}_2, \mathbf{x}_1 ).
$$
We integrate by parts, assuming that the integration region is such that
no boundary contributions appear and the relations (\ref{LT}, \ref{Lprojt}) 
apply.
$$ E(\mathbf{u}\p)\int d\mathbf{x}\p_1 d\mathbf{x}\p_2 
 (L^{-1}_{1\p}(u_1^{\prime +}, u_1\p))^T (L^{-1}_{2\p}(u_2^{\prime +},
u_2\p))^T \psi(\mathbf{x}\p_1, \mathbf{x}\p_2)
\Phi(\mathbf{x}\p_2, \mathbf{x}\p_1,\mathbf{x}_2, \mathbf{x}_1 )
= $$ $$
\frac{E(\mathbf{u}\p)}{u_1\p (u_1^{\prime +} +1) u_2\p (u_2^{\prime +} +1)}
\int d\mathbf{x}\p_1 d\mathbf{x}\p_2
L_{1\p}(u_1\p -n, u_1^{\prime +}) L_{2\p}(u_2\p -n, u_2^{\prime +})
\psi(\mathbf{x}\p_1, \mathbf{x}\p_2)
\Phi(\mathbf{x}\p_2, \mathbf{x}\p_1,\mathbf{x}_2, \mathbf{x}_1 ).
$$
The relation proven for arbitrary functions implies the operator relation
(\ref{QLL}).

If $\Phi_{2\p 1\p 2 1}$ in (\ref{defQ} ) is constructed by $R$ operations from 
$\Phi_{2\p 1\p 2 1; 0}  = \delta (\mathbf{x}\p_1) \delta (\mathbf{x}\p_2)$
then its parameters $\mathbf{v}$ are permutations of
$$ \mathbf{v}^{(0)} =
\begin{pmatrix}
2\p & 1\p & 2& 1 \\
2^{\prime +}& 1^{\prime +} & 2 & 1 
\end{pmatrix}.
$$ and $E(\mathbf{u}) = E(\mathbf{v}^{(0)}) = v_1\p v_2\p (v_1+1) (v_2+1) $.

\subsubsection{ The point permutation operator}
\noindent
In the case of $ \Delta = \Phi_{2\p 1\p 2 1; 2}^{--++}$ 
(\ref{Phi4--++R}) we have
$$ \Delta (\mathbf{x}_2\p, \mathbf{x}_1\p, \mathbf{x}_2, \mathbf{x}_1)
= R_{22\p}(v_{1\p}^+ -v_2) R_{12\p}(v_{2\p}^+ -v_1) 
\delta (\mathbf{x}\p_1) \delta (\mathbf{x}\p_2),  $$
$$  \mathbf{v}^{(2)}  =
\begin{pmatrix}
2\p&1\p&2&1 \\
1&2&1\p&2\p   
\end{pmatrix}. $$

$\mathbf{u}\p = \mathbf{v}^{(2)}$ means 
$ v_{1\p} =  u_{1\p}, v_{2\p} = u_{2\p}, v_1 = u_1, v_2 = u_2 $, 
$ u_{2\p}^{ +} = v_1, u_{1\p}^{ +} = v_2,  
u_2^{ +} = {v_{2\p}}\p , u_1^{ +} = {v_{2\p}}^+ $
and  this implies $F(\mathbf{u}\p) = 1$. 
We obtain as the specification of (\ref{QLL}) with the kernel in (\ref{defQ})
substituted by $\Delta$ and 
$\mathbf{u}\p = \mathbf{v}^{(2)} $
that the corresponding operator $\hat \Delta$
defined with this kernel obeys 
\be \label{P12}
 \hat \Delta L_1(v_{1\p}-n, v_2) L_2(v_{2\p}-n, v_1) 
 = L_2(v_{1\p}-n, v_2) L_1(v_{2\p}-n, v_1) \hat \Delta.
\ee
Thus choosing $\Delta = \Phi_{2\p1\p 21;2}^{--++}$ 
(\ref{Phi4--++R}) as the kernel in (\ref{defQ})
 we obtain the representation of the operator of permutation $P_{12}$ acting as
$ P_{12} ({x}_{1, a}, \dd_{1,a}) P_{12} = ({x}_{2,a} , \dd_{2, a})$.

\vspace{.5cm}

\subsubsection{ The parameter pair permutation operators} 
\noindent
We show that the kernel $R^{++}_{1\p 2\p}(v_1-v_2)  \Delta 
= \Phi^{--++}_{2\p 1\p 2 1;3 }$ defines by (\ref{defQ})
the operator $\hat R_{12} = P_{12} R_{12}$, where $R_{12} $ obeys 
(\ref{RLLproj1}). 
Indeed, the parameter permutation pattern  is
$$ \mathbf{v}^{(3)} = 
\begin{pmatrix}
2\p&1\p&2&1 \\
2&1&{1\p}^+&{2\p}^+
\end{pmatrix}. $$
 We see that $F(\mathbf{u}\p) = 1$ and the relation (\ref{QLL}) specifies with
$\mathbf{u}\p = \mathbf{v}^{(3)} $ to
\be \label{R12I}
\hat R_{12}(v_1-v_2) L_1 (v_{1\p}^+, v_1)  L_2(v_{1\p}^+, v_2)
= L_2(v_{1\p}^+, v_2) L_1(v_{2\p}^+, v_1) \hat R_{12}(v_1-v_2). 
\ee
The relation is to be compared with (\ref{RLLproj1}).
The parameters $v_1, v_2$ and the underlying canonical pairs are permuted.
and the  parameters $\tilde v_1^+ = v_{2\p}^+ $ and $\tilde v_2^+ =
v_{1\p}^+$ remain.

Next we show that the kernel $R_{2\p 1\p}(v_{2\p} - v_{1\p}) \Delta $ 
defines by (\ref{defQ})
the operator $\hat R_{21} = P_{12} R_{21}$, where $ R_{21}$ 
obeys (\ref{RLLproj2}). 
Indeed, the permutation pattern  is
$$ \mathbf{v}^{(3\p)} =
\begin{pmatrix}
1\p&2\p&2&1 \\
1&2&{1\p}^+&{2\p}^+
\end{pmatrix}. $$
Again $F(\mathbf{u}\p) = 1$. The relation (\ref{QLL}) specifies with
$\mathbf{u}\p = \mathbf{v}^{(3\p)}$ to
\be \label{R21I}
 \hat R_{21}(v_{2\p} - v_{1\p})  L_1 (v_{2\p}^+ , v_2)  L_2(v_{2\p}^+,
v_1)
=L_2(v_{1\p}^+, v_2) L_1(v_{2\p}^+, v_1) \hat R_{21}(v_{2\p} - v_{1\p}). 
 \ee
This relation is to be compared with (\ref{RLLproj2}).
The parameters $\tilde v_1^+ = v_{2\p}^+, \tilde v_2^+ = v_{1\p}^+$ are
permuted together with the underlying canonical pairs and the parameters 
$v_1, v_2$ remain.

\vspace{.5cm}

\subsubsection{ The complete Yang-Baxter operator}  
\noindent
The kernel $ R_{2\p 1\p}(v_{2\p} - v_{1\p}) R_{1\p 2\p}(v_1-v_2) 
\Delta= \Phi^{--++}_{2\p1\p21;4}$  (\ref{Phi4--++R})
defines by (\ref{defQ})
the  complete Yang-Baxter operator $\hat \R_{12}$. 
Indeed, the permutation pattern of parameters is   
$$ \mathbf{v}^{(4)} =
\begin{pmatrix}
1\p&2\p&2&1 \\
2&1&{1\p}^+&{2\p}^+
\end{pmatrix}. $$
We find $F(\mathbf{u}\p) = 1$ and the relation (\ref{QLL}) results with 
$\mathbf{u}\p = \mathbf{v}^{(4)} $
in
\be \label{YBnormal}
 \hat \R_{12} (\tilde v_1^+- \tilde v_2^+, v_1-v_2) L_1(\tilde v_1^+, v_1)  
L_2(\tilde v_2^+, v_2)
= L_2(\tilde v_2^+, v_2) L_1(\tilde v_1^+, v_1) 
\hat \R_{12}(\tilde v_1^+- \tilde v_2^+, v_1-v_2). 
 \ee
where $\tilde v_1^+ = v_{2\p} -n,  \tilde v_2^+ = v_{1\p} -n $.
Both parameters and the  underlying canonical pairs are permuted.
The weights corresponding to the points of the correlator
 $ R_{2\p 1\p}(v_{2\p} - v_{1\p}) R_{1\p 2\p}(v_1-v_2) 
\Delta = \Phi^{--++}_{2\p1\p21;4}$  are calculated from
$\mathbf{v}^{(4)}$ as
$ 2\ell_{2\p} = - 2\ell_2-n = v_2-v_{1\p} $, $2\ell_{1\p} = -2\ell_1 -n = v_1
-v_{2\p} $.
The weights in the $L$ operators $L_1$ and $L_2$ are
$\tilde v_1^+ - v_1 = 2\ell_1 $ and $\tilde v_2^+ -v_2 = 2\ell_2 $.
The operator $\R_{12}$ depends on these weights and the 
spectral parameter difference $ v_1-v_2 $.

\subsubsection{ The operators of the crossed YSC}

We have constructed a completely connected correlator by crossing the 
former one (\ref{Phi4c}), 
$ R_{12}(v_{1\p}^+ - v_{2\p}^+) R_{23} (v_2-v_1)  
 \Phi^{--++}_{2\p1\p21;4}$. 
Its parameter pattern is obtained from $
\mathbf{v}^{(4)}$ (\ref{v4}) by permuting the last two columns and
becomes (\ref{v4c}), 
$$ \mathbf{v}^{(4) c} =
\begin{pmatrix}
1\p&2\p&1&2 \\
2&1&{2\p}^+&{1\p}^+
\end{pmatrix}. $$
Whereas in $ \Phi^{--++}_{2\p1\p21;4}$ the weights of the
next-to nearest points are related, $2\ell_{2\p} + 2\ell_2 +n =0$,
we have here  $2\ell_{2\p} + 2\ell_1 +n =0$. 
Now  the relation (\ref{QLL}) results with 
$\mathbf{u}\p = \mathbf{v}^{(4) c} $ in
\be \label{Q}
 \Q L_1(v_{2\p}^+, v_1) L_2(v_{1\p}^+ - v_2) = L_2 (v_{2\p}^+, v_1) L_1 (v_{1\p}^+ - v_2)
\Q \ee
It compares with the relation (\ref{P12}) for $\hat \Delta $. Both act on the 
monodromy $L_1 L_2$ like the permutation of points $P_{12}$, but $\Q$
is depends not only on the weights but additionally on $v_1-v_2 $.

The YSC obtained from the latter by the cyclic permutation has the
property that the weights of the adjacent points are related, because now
the sequence of points is $2\p, 1, 2, 1\p $. The parameter permutation
pattern is
$$ \mathbf{v}^{(4) c c} =
\begin{pmatrix}
2\p&1&2& 1\p-n \\
1&{2\p}^+&{1\p}^+& 2-n
\end{pmatrix}. $$
Now  the relation (\ref{QLL}) results with 
$\mathbf{u}\p = \mathbf{v}^{(4) c c} $ in 
\be \label{Qc} 
\Q^c L_1( v_1^+, v_{2\p}^+) L_2 (v_{2\p}^+, v_1) = L_2(v_{1\p}^+ , v_2)
L_1(v_2^+, v_{1\p}^+) \Q^c .\ee
The weights of the left monodromy are related and the weights of the right
monodromy are related as well, but the weights of $L_1, L_2$ on the left 
hand side are not
related to the weights appearing on the right. Unlike the previous cases
the parameter arguments of the $L$ operators on the left hand side are not
permutations of the ones on the right hand side.  
The operator
depends besides of the weights additionally on $v_1-v_2$.

\section{Application to scattering amplitudes}
 \setcounter{equation}{0}

\subsection{The three-point YSC and the three-point amplitudes}

We consider the completely connected 3-point YSC in the helicity 
form (\ref{linkhel}), in particular (\ref{Phi3-++}). 

$$ \Phi_{123,2}^{\lambda -++} = \int \frac{dc_{13} dc_{12}}{c_{13}^{1+u_1^+-u_3}
c_{12}^{1+u_3-u_2} } \delta^{(m)} (\bar \lambda_1 - c_{12} \bar \lambda_2 -
c_{13} \bar \lambda_3 ) 
\delta^{(m)} (\lambda_2 + c_{12} \lambda_1) \delta^{(m)} (\lambda_3 + c_{13}
\lambda_3) ) $$
The monodromy parameters are
$$ \begin{pmatrix}
1 & 2 & 3 \\
2 & 3 & 1^+
\end{pmatrix}
$$
and result in the weights
\be \label{l1l2l3}
 2\ell_1 = u_2-u_1, 2\ell_2 = u_3-u_1 , 2\ell_3 = u_1^+ -u_3. \ee 
The conditions in the delta arguments  lead to
\be \label{mom}
  \lambda^1_{\alpha} \bar \lambda^1_{\alphadot} + \lambda^2_{\alpha} \bar
\lambda^2_{\alphadot}
+ \lambda^3_{\alpha} \bar \lambda^3_{\alphadot}= 0 . \ee
Indeed, 
multiply the condition in the first delta by $\lambda^1_{\alpha}$,
$$
 \bar \lambda^1_{\alphadot} \lambda^1_{\alpha}
  - c_{12} \bar \lambda^2_{\alphadot} \lambda^1_{\alpha} 
-c_{13} \bar \lambda^3_{\alphadot} \lambda^1_{\alpha} = 0  
$$
We use the other deltas to substitute
$  c_{12} \lambda^1_{\alpha} = \lambda^2_{\alpha}, c_{13} \lambda^1_{\alpha}
= \lambda^3_{\alpha} $
and obtain the momentum condition (\ref{mom}).

One non-trivial projection of the second and third conditions are sufficient
for fixing the integration variables:
$$ c^*_{12} = - \frac{<2 \mu> }{<1 \mu>}, 
\ c^*_{13} = - \frac{<3 \mu\p> }{<1 \mu\p>}.
$$
We may choose $\mu = |3> $ and  $\mu\p = |2> $ .
The result of the  integration over $c_{12}, c_{13}$ is 
$$  \left (\frac{<1 \mu\p> }{<3 \mu\p>}\right )^{u_1^+ -u_3} 
\left (\frac{<1 \mu> }{<2 \mu>}\right )^{u_3 -u_2} 
$$
up to Jacobi factors not depending on the parameters. 
The latter can be fixed by comparing the scaling of the expression 
under $\lambda_I \to \mu_I^{-1} \lambda_I $
with the 
scaling law of the YSC in the helicity form, $\mu_I^{2h_I}$.

For $m=2$ we obtain, using  (\ref{l1l2l3}) and (\ref{2h2l}),
$$\Phi^{\lambda -++} =
 <12>^{2 h_3+ 1}   <13>^{2 h_2+ 1}   <23>^{2 h_1+ 1}
\delta^{(4)} ( \bar \lambda^1 \lambda^1
 + \bar \lambda^2 \lambda^2 + \bar  \lambda^3 \lambda^3 ) 
$$

The expression (\ref{M3h})  is reproduced for the case $\eta=+1$.
The case $\eta=-1$ is obtained in the analogous way from 
$\Phi_{123,2}^{--+}$ (\ref{Phi3--+}). 

Thus we have shown that the 3-point amplitudes are 3-point YSC.

\subsection{ The convolution construction and BCFW }

In sect. 3.2.1 and 3.5.2 we have shown in the particular case of three points 
that the symmetric convolution of YSC allows to construct further YSC.
The generalization to higher point YSC is obvious. 
The convolution has been formulated in terms of the canonical pairs
$x_a, \dd_a, a= 1, ..., n$, in particular the convolution 
means integration over the coordinates $x_a$ of an identified point of the
two correlators. The convolution construction can be reformulated
for YSC in other representation forms, in particular in the helicity
form if $n=2m$. Then the integration is over the helicity coordinates
$ \lambda_{\alpha}, \bar \lambda_{\alphadot}, 
\alpha = 1, ..., m, \alphadot = 1, ..., m$ of the identified point.
In the application to scattering amplitudes of QCD ($n= 2m = 4$)
this integration can be
rewritten in terms of the on-shell momentum components of the fused legs
$ P_{\alpha \alphadot} = \lambda_{\alpha} \ \lambda_{\alphadot}$, 
 $d^4 P \delta (P^2) $  and a phase. 
Thus we obtain a form compatible with the second factor in (\ref{BCFW}).

The energy-momentum delta distribution included in each correlator
results in the overall 
energy-momentum delta distribution of the resulting correlator. 

 In sect. 3.5.2 we have seen that the result 
of the convolution of two completely connected 
3-point YSC is a 4-point YSC, but not a completely connected one.
A particular $R$ operation (\ref{fusion33}) 
leads to the completely connected YSC.
The $R$ operator can be expressed by a contour integral over a shift
operator (\ref{R12}). In the helicity representation it appears 
by (\ref{x1d2}) as
$$ R_{IJ}(u) = \int_{\mathcal{C}} \frac{dc}{c^{1+u} } 
\exp(- c \bar \lambda_{J \alphadot} \bar \dd_I^{\alphadot}  
+c \lambda_{I \alpha} \dd_J^{\alpha} ).
$$
We observe that the first factor in (\ref{BCFW}) is a particular
case of the latter $R$ operator expression.

From this comparison we conclude that the convolution construction
of YSC supplemented by $R$ operations generalizes the 
BCFW construction of QCD tree amplitudes.

\subsection{4-point YSC and parton tree amplitudes}

We consider the four-point YSC at $n=4$ with $K=2$ in the helicity
representation 

\be
\Phi^{--++ \lambda} = \int \mathrm{d}^4c \, \varphi (c)\, 
\delta^{(2)}(\bar\lambda_1-c_{13}\bar \lambda_3-c_{14}\bar\lambda_4) 
\delta^{(2)}(\bar \lambda_2-c_{23}\bar \lambda_3-c_{24} \bar\lambda_4) 
\delta^{(2)}(\lambda_3 +c_{13}\lambda_1 + c_{23}\lambda_2) \ee $$
\times \delta^{(2)}(\lambda_4 +c_{14}\lambda_1 + c_{24}\lambda_2). 
$$
and do the integrals over
the $c$ variables with the result 
\be \label{philambda}
\Phi^{--++ \lambda} = \varphi(c^*) \ \ \delta^{(4)}(\sum k_i), \ee
$$ c_{13}^* = \frac{[14]}{[34]}, \  
 c_{14}^* = \frac{[13]}{[43]}, \ 
 c_{23}^* = \frac{[24]}{[34]}, \   
 c_{24}^* = \frac{[23]}{[43]}, \ 
c_{13}^* c_{24}^*-  c_{14}^* c_{23}^*= \frac{[12]}{[34]}. $$  
Here we denote $ (k_i)_{\alphadot, \alpha} = 
\bar \lambda_{i,\alphadot} \lambda_{i,\alpha}$,   
$[ij] = \bar \lambda_{i,1} \bar \lambda_{j,2} - \bar \lambda_{i,2} \bar
\lambda_{j,1}$.

The gluon and quark helicities $h_i$ are related to the weights as
$ 2h_i = 2\ell_i +2 $.
The YSC $\Phi_X$ (\ref{Phi4--++v}) and $\Phi_{||}$ (\ref{Phi4c}) 
depend on two independent helicities  and one extra
parameter $\e= v_3-v_4$. In the case of $\Phi_X$ the helicities are related by
$ h_1 = - h_3, \ \ h_2 = -h_4 $.
With these conditions only 4 of the 6 helicity configurations of the
$2 \to 2$ helicity amplitudes are accessible.  
With $\Phi_{||}$ we
 cover the remaining cases. Here the helicities are
related by
$ h_1 = - h_4, \ \ h_2 = -h_3 $.
With these conditions again 4 helicity configuration cases can be covered,
two of them doubling some of the above correlator.

From (\ref{philambda}) with $\varphi(c)$ from (\ref{Phi4--++v}) and
(\ref{Phi4c})
we obtain the explicit expressions as functions of the
independent helicities and the extra parameter $\e$, 
$$ \varphi_{X}(c^*; h_1,h_2, \e) = 
\left ( \frac{ [14] [23]}{[12][34]} \right )^{\e} \
\frac{[12]^{1+2h_1} [34]^{1-2h_2} }{ [14] [23]^{1+2h_1-2h_2} },
$$ 

$$ \varphi_{||}(c^*; h_1,h_2, \e) = 
\left ( \frac{ [13] [24]}{[12][34]} \right )^{\e} \
\frac{[12]^{1+2h_1} [34]^{1-2h_2} }{ [14] [23] [24]^{2h_1-2h_2} }.
$$

We observe that the gluon tree amplitudes in spinor-helicity form
\cite{Dixon96,Khoze04}
are reproduced by these two
YSC as

$$ M(1,1,-1,-1) = \varphi_{X}(c^*, 1,1,0), \ \    
M(-1,-1,1,1) = \varphi_{X}(c^*, -1,-1,0), \ \
$$ $$
 M(1,-1,-1,1) = \varphi_{X}(c^*, 1,-1,4), 
 M(-1,1,1,-1) = \varphi_{X}(c^*, -1,1,0), $$

$$ M(1,1,-1,-1) = \varphi_{||}(c^*, 1,1,0), \ \    
M(-1,-1,1,1) = \varphi_{||}(c^*, -1,-1,0), \ \
$$ $$ 
M(1,-1,1,-1) = \varphi_{||}(c^*, 1,-1,4), 
 M(-1,1,-1,1) = \varphi_{||}(c^*, -1,1,0).
$$

The quark tree amplitudes are reproduced as

$$
M(\half, \half,-\half ,-\half) = \varphi_X(c^*, \half,\half, 0), $$ $$ 
M(\half, -\half,-\half, \half) = \varphi_X(c^*, \half,-\half, 3), \ \ 
M(-\half, \half,\half, -\half) = \varphi_X(c^*, -\half,\half, 1), \ \ 
$$
$$
M(\half, \half,-\half ,-\half) = \varphi_{||}(c^*, \half,\half, 0), $$ $$ 
M(\half, -\half,\half, -\half) = \varphi_{||}(c^*, \half,-\half, 3), \ \ 
M(-\half, \half,-\half, \half) = \varphi_{||}(c^*, -\half,\half, 1).  
$$

The gluon-quark tree amplitudes are reproduced as

$$ M(1,\half, -1,-\half) = \varphi_X(c^*, 1, \half, 1), \ \ 
M(1,-\half, -1,\half) = \varphi_X(c^*, 1, -\half, 3),
$$
$$
M(1,\half, -\half,-1) = \varphi_{||}(c^*, 1, \half, 1), \ \ 
M(1,-\half, \half,-1) = \varphi_{||}(c^*, 1, -\half, 3).
$$

To obtain the tree amplitudes the  monodromy parameters of the YSC 
are fixed by the values of the helicities besides of the parameter $\e$.
The physical criterion for choosing $\e$  is to allow only single poles in 
$s=|<12>|^2 = |<34>|^2, t= |<14>|^2 =|<23>|^2 $ or $u= |<13>|^2 = |<24>|^2
$.

\section{Application to the Bjorken asymptotics}

\subsection{Three-point YSC and the parton splitting amplitude}

We consider the 3-point YSC (\ref{Phi3-++}), (\ref{Phi3--+})
in the case $n=2, m=1$
converted to the helicity form according to (\ref{linkhel}).
We start with (\ref{Phi3-++}).
$$ \Phi_{123,2}^{-++, \lambda} = \int \frac{dc_{13} dc_{12} }{c_{13}^{1+u_1^+ - u_3}
c_{12}^{1+u_3-u_2} } 
\delta (\bar \lambda_1 - c_{12} \bar \lambda_2 -c_{13} \bar \lambda_3)
\delta (\lambda_2 + c_{12} \lambda_1) \delta (\lambda_3 + c_{13} \lambda_1). 
$$
The delta distributions fix the values of $c_{12}, c_{13}$.
$$ \Phi_{123,2}^{-++, \lambda} = 
\lambda_1^{1+u_1^+ -u_2} \lambda_2^{-1-u_3+u-2} \lambda_3^{-1 -u_1+u_3}
\delta(\lambda_1 \bar\lambda_1 + \lambda_2 \bar\lambda_2 + \lambda_3
\bar\lambda_3 ). 
$$
From the monodromy parameters (\ref{u3-++}) we obtain the 
weights as $ 2\ell_1 = u_2-u_1, 2\ell_2= u_3-u_2, 2\ell_2 = u_1^+ - u_3$.
We recall  the relation to the helicity (\ref{2h2l}), (\ref{2h2l1}),
$2h = 1+ 2\ell$,  and express the exponents in terms of helicities
constrained by $2h_1+2h_2+2h_3 = +1$,

$$ \Phi_{123,2}^{-++, \lambda} = \lambda_1^{-h_1} \lambda_2^{-h_2}\lambda_3^{-h_3} 
\delta(\lambda_1 \bar\lambda_1 + \lambda_2 \bar\lambda_2 + \lambda_3
\bar\lambda_3 ). $$

We treat (\ref{Phi3--+}) in analogy
$$ \Phi_{123,2}^{--+, \lambda} = \int \frac{dc_{13} dc_{23} }{c_{13}^{1+u_1^+ - u_3}
c_{23}^{1+u_2^+-u_1^+} }
\delta(\bar \lambda_1 - c_{13} \bar \lambda_3) 
\delta(\bar \lambda_1 - c_{12} \bar \lambda_2)
\delta(\lambda_3 +  + c_{13}  \lambda_1 +c_{23} \lambda_2)
$$
and obtain
$$ \Phi_{123,2}^{--+, \lambda}  =
\bar \lambda_1^{h_1} \bar \lambda_2^{h_2}\bar \lambda_3^{h_3} 
\delta(\lambda_1 \bar\lambda_1 + \lambda_2 \bar\lambda_2 + \lambda_3
\bar\lambda_3 ). $$
The monodromy parameters of this correlator lead to the constraint on the
weights $2\ell_1 + 2\ell_2 + 2\ell_3 +4 = 0$ and therefore the helicity
parameters obey $2h_1 + 2h_2+ 2 h_3 = -1$.

If  $k_I = \lambda_I \bar \lambda_I $ represent light-cone components of
momenta and therefore take real values,
 we rewrite both results by separating the phase
 and the momentum conservation factors as
$$ \Phi_{123,2}^{ \lambda} =
 e^{i\alpha}  M (k_1, k_2, k_3, h_1,h_2,h_3) 
\delta(k_1+k_2+k_3). $$
With the notation 
\be \label{2h1}
2h_1 + 2h_2+ 2h_3 = \eta, \ \ \ \eta=\pm 1 \ee
we write for both cases
$$ M(k_1, k_2, k_3, h_1,h_2,h_3) = k_1^{-h_1 \eta} k_2^{-h_2 \eta} 
k_3^{-h_3 \eta} . 
$$
We define the collinear parton triple vertex differing by
 a factor related to the propagators in the convolutions.
\be \label{M3}
 M_{3}(k_1,k_2,k_3; h_1,h_2,h_3) =  (k_1k_2k_3 )^{\half}
M(k_1, k_2, k_3, h_1,h_2,h_3) = k_1^{\half -h_1 \eta} k_2^{\half-h_2 \eta} 
k_3^{\half-h_3 \eta}.
\ee
In the case of the scattering amplitudes, sect. 4,  
the helicity parameters in the related YSC are identified
with the parton helicities 
 (up to sign for the all-ingoing convention). 
 There the YSC helicities obey  
in the case of 3 points $2 h_1 + 2h_3 + 2h_3= 2\eta $, because $n=2m= 4$.
Here $n= 2m=2$ and therefore the   helicity parameters obey (\ref{2h1}). 

We observe that the collinear parton triple vertex 
coincides with  the parton splitting amplitude (\ref{M3split}), 
resulting in the
parton splitting probability  (\ref{WMsplit}) $A\to B C$ by 
substituting for $B, C$ the momentum fractions $z, 1-z $ 
 and
the physical parton helicities with signs regarding the all-ingoing convention. 
$$ W_{A BC}(z) = |M_3(1, -z, z-1; h_B + h_B-\half \eta, -h_B, -h_C) |^2 .
$$
For comparison we write some results for the splitting kernels
(up to contributions proportional to $\delta(1-z)$) as calculated 
from the splitting amplitude by
(\ref{WMsplit}). Recall that the labels at $W_{AB}$ refer to the
physical helicities of the involved quarks ($\pm \half$) 
and gluons ($\pm 1$). 
$$ W_{-1,-1}(z) = W_{+1, +1}(z) = W_{+1,+1,-1}(z) + W_{+1,+1,+1}(z) =
\frac{1}{z (1-z) } +  \frac{z^3}{(1-z)}, $$
\be \label{W111}
 W_{-\half, -\half} (z) = W_{+\half, +\half} (z) = 
\frac{z}{1-z}, \ee
$$ W_{-1, 1} = W_{+1, -1} (z) =  \frac{(1-z)^3}{z}, \ \ \  
 W_{1, -\half} (z) = (1-z)^2,  \ \ \ W_{1, +\half} (z) = z^2. $$

\subsection{Four-point YSC and $RLL$ relations }

We have shown in sec. 3.6  that four-point YSC define as kernels operators 
obeying RLL type relations involving the products of two $L$ operators.
In particular, the integral operator with the YSC (\ref{Phi4--++v})
obeys the Yang-Baxter relation (\ref{RLL}), (\ref{YBnormal}). 
Here we specify the $L$ operators to the case $n=2m=2$ and use properties
derived from the explicit form in normal coordinates.

The reduction of the $L$ operators to the subspace of eigenvalue
$2\ell$  is done  for the  action  on functions of
the coordinate components by the 
restriction to a definite degree of homogeneity.
We specify (\ref{x2l}), (\ref{projL}) to the present case.
$$
 \psi_{2\ell} (\mathbf{x}) =( x_2)^{2\ell}  \phi(x), \ \
L^+(u) ( x_2)^{2\ell} \phi(x) = ( x_2)^{2\ell}   L_x(u^+ +1,u+1) \phi(x), $$
\be\label{Lred}  \mathbf{x} = (x_1,x_2), \ \ \ x = \frac{x_1}{x_2} . 
\ee
We change to the canonical pair $x, \dd$, where $x$ is the normal 
coordinate ratio and $\dd$ the corresponding derivative operator,
using 
$ x_1 \dd_1 \to x \dd, \ 
x_2 \dd_2 = (\mathbf{x}\mathbf{p}) - x_1 \dd_1 \to 2\ell - x \dd $.
Explicitly we have
$$ L^+(u) =
\begin{pmatrix}
  u+1 + x_1 \dd_1 & x_2 \dd_1 \\
x_1 \dd_2 & u+1 + x_2 \dd_2 
\end{pmatrix}, \ $$  
$$
  L_x(u^++1, u+1) =
\begin{pmatrix}
  u+1 + x \dd &  \dd  \\
x (- x \dd + 2\ell)  & u+1 + 2\ell - x \dd 
\end{pmatrix},
$$
where $u^+ = u+2\ell $ as above.  
A more symmetric definition of the
parameters is to substitute  $v= u+\ell $ and to write 
\be \label{Lv1v2}
 L_x(u^+, u) = 
 L(v^{(1)}, v^{(2)}), \ \  
 v^{(1)} = v + \ell, v^{(2)} = v-\ell-1. \ee
The result for $L$ compares with (\ref{Mabx}),
$L_{ab} (v^{(1)},v^{(2)}) = v \delta_{ab} + M_{ab}$.  
The $L$ matrix and the $s\ell(2)$ generators are related as
$$ L(v^{(1)},v^{(2)}) = 
\begin{pmatrix}
v + S^0 & S^- \\
S^+ & v - S^0 
\end{pmatrix}, 
$$
\be \label{Sa}
 S^- = \dd, \ \ S^+ = x(-x\dd + 2 \ell),  \ \ S^0 = x\dd -\ell. \ee

We continue with the formulation in terms of the normal coordinates.
Considering  more than one point we label now by the index at $x$
the point $I$, i.e. in the following $x_1, x_2$ are not the components
 of $\mathbf{x}$ as in (\ref{Lred}).
 Let $\psi(x_1,x_2)$ obey the eigenvalue relation 
$$ \mathrm{R}_{21}(w) \psi(x_1,x_2) = \lambda(w) \psi(x_1,x_2). $$
 The Yang-Baxter  relation (\ref{RLL}) implies
\be \label{Qv} 
\mathrm{R}_{21}(w) L_{2}(v) L_{1}(v-w) \psi(x_1,x_2) = \lambda (w)  
 L_{1}(v-w) L_2(v)  \psi(x_1,x_2).
\ee
We recall the known fact that the RLL relation implies the eigenvalues of
the operator $R_{21} $. For this we do not use its
integral operator form, which will be written explicitly in the next
subsection.  

We  abbreviate the relation (\ref{Qv}) as 
$ \mathrm{R} \mathcal{L} \psi= \lambda \mathcal{R} \psi $ 
and expand in powers of $v$. At $v^2$ we have a trivial relation, at $v^1$
we find:  
If $\psi$ is an eigenfunction then it is also $(S_1^a + S_2^a) \psi$ 
with the same eigenvalue, 
i.e. we have $s\ell(2)$ irreducible representation subspaces  
of degeneracy.
The conditions at $v^0$ are related by this global symmetry. Therefore it is
sufficient to analyze the condition related to one matrix element, e.g.
$12$. 
 $$
\mathcal{R}_{12}= (S_1^0 -\delta) S_2^- - S_1^-S_2^0, 
\qquad \mathcal{L}_{12}= S_2^0 S_1^- - S_2^-(S_1^0 + \delta). $$
We consider the function
$\psi_n^{(0)} = (x_1-x_2)^n $ 
being a lowest weight vector
$$ (S_1^- +S_2^-) \psi_n^{(0)} = 0, \quad (S_1^0 +S_2^0) \psi_n^{(0)} = - \mu_n
\psi_n^{(0)}, \quad \mu_n = -\ell_1- \ell_2 +n. $$
We calculate  
$$ \mathcal{R}_{12} \psi_n^{(0)} =  
 \half \mu_n \mathrm{S}^- \psi_n^{(0)} + \half (1+\delta) \mathrm{S}^-
\psi_n^{(0)}, \quad 
 \mathcal{L}_{12} \psi_n^{(0)} = 
-\half \mu_n \mathrm{S}^- \psi_n^{(0)} - \half (1-\delta) \mathrm{S}^-
\psi_n^{(0)}, $$
where $\mathrm{S}^a= S_1^a - S_2^a$.
We obtain by (\ref{Qv})
$$ \lambda_n [\mu_n + 1 + \delta] \, \mathrm{S}^- \psi_n^{(0)} =  
[ - \mu_n + \delta -1] \, \mathrm{R}_{21} \, \mathrm{S}^- \psi_n^{(0)}. $$
We have marked the eigenvalue of the representation generated from the
lowest weight vector $\psi_n^{(0)}$ by the index $n$. 
We find that also $\mathrm{S} ^- \psi_n^{(0)}$ is an eigenfunction of
$\mathrm{R}_{21}$
with another eigenvalue,
$$ \mathrm{R}_{21} \ \mathrm{S} ^- \psi_n^{(0)} = \lambda_{n-1} \mathrm{S} ^-
\psi_n^{(0)}. $$
By comparison we obtain a recurrence relation for $\lambda_n$.
It is easy to see that the constant function is an eigenfunction and moreover
a lowest weight vector. Thus as the result of the analysis 
we  find all eigenfunctions and eigenvalues. The latter obey the iterative
relation

$$\lambda_{n+1} = - \lambda_n  \frac{n-\ell_1-\ell_2+w}{
n-\ell_1-\ell_2-w } $$
solved by 
\be \label{REW}
 \lambda_n(w) = (-1)^n \frac{\Gamma(n-\ell_1-\ell_2+w) }{ 
\Gamma(n-\ell_1-\ell_2-w) } 
\frac{\Gamma(-\ell_1-\ell_2-w) }{ \Gamma(-\ell_1-\ell_2+w) }. 
\ee

\subsection{Four-point YSC and evolution kernels in positions}

We specify the general integral form of YSC (\ref{link}) to
$N=4, K=2, n=2m=2$.
\be \label{link1}
 \Phi^{--++} = \int \mathrm{d}^4c \, \varphi (c) \,
\delta^{(2)}(\mathbf{x}_1 -c_{13}\mathbf{x}_3 - c_{14} \mathbf{x}_4) 
\delta^{(2)}(\mathbf{x}_2 -c_{23}\mathbf{x}_3 - c_{24} \mathbf{x}_4),  
\ee 

We have four delta distributions.
The vanishing of the arguments of the first two results in
$$ c_{13}^{(0)} = \frac{\langle 14 \rangle }{\langle 34 \rangle}, \ \ \  c_{14}^{(0)} =
\frac{\langle 13 \rangle}{ \langle 43 \rangle}. $$
We use the notations 
$\langle IJ \rangle  = x_{I,1} x_{J,2} - x_{I,2} x_{J,1}, \ \ x_I =
\frac{x_{I,1} }{x_{I,2}} $.
 The last two result in
$$ c_{23}^{(0)} = \frac{\langle 24 \rangle}{\langle 34 \rangle}, \ \ \  c_{24}^{(0)} =
\frac{\langle 23 \rangle}{\langle 43 \rangle}.$$
Further,
$$c^{(0)}_{13} c^{(0)}_{24} - c^{(0)}_{14} c^{(0)}_{23} = \frac{\langle 12 \rangle}{\langle 34 \rangle}. $$
The Jacobi determinant is
$$ \frac{\dd (x_{1,1}\p, x_{1,2}\p, x_{2,1}\p, x_{2,2}\p )}
{\dd (c_{13}, c_{14},c_{23}, c_{24}) } = \langle 34\rangle^2. $$
$x_{1,1}\p, x_{1,2}\p, x_{2,1}\p, x_{2,2}\p $ abbreviate the arguments of
the delta distributions in (\ref{link1}). 

The result is 
\be \label{Phi1234} 
 \Phi_{1234}^{--++} = \langle 34\rangle^{-2} \varphi^{--++}(c^{(0)}). 
\ee
We change from the homogeneous to the normal coordinates
as in the previous subsection,
$$ \langle IJ \rangle  = x_{I,2} x_{J,2} x_{IJ} , \ \ x_{IJ} = x_I - x_J,  $$
and extract the scale factor 
\be \label{phi}
 \Phi_{1234} = x_{1,2}^{2\ell_1} x_{2,2}^{2\ell_2}x_{3,2}^{2\ell_3}
x_{4,2}^{2\ell_4}  \phi_{1234}. \ee
$ \phi_{1234}$ is obtained from  $\Phi_{1234} $ by 
replacing $\langle IJ \rangle$ by $x_{IJ}$.

In the case of the  4-point YSC ( \ref{Phi4--++v})
we obtain by (\ref{Phi1234}), (\ref{phi})
$$\phi_X (x_1,x_2,x_3,x_4)= \frac{x_{34}^{1+v_1^+ -v_3} }{ x_{23}^{1+v_1-v_2}
x_{14}^{1+v_4-v_3} x_{12}^{1+v_2^+-v_4}} =
\frac{x_{12}^{1+2\ell_1-\e} }{x_{23}^{1+2\ell_1-2\ell_2-\e} x_{14}^{1-\e}
x_{24}^{1+2\ell_2+ \e} }$$
where $2\ell_1= v_3-v_1, 2\ell_2 = v_4-v_1, \e= v_3-v_4$. 
We substitute $1,2,3,4,\to 2\p ,1\p,2,1$  and obtain the explicit form of
the kernel of the $R$ operator $R(\e) $ 
obeying the RLL relation (\ref{RLL}), (\ref{YBnormal}) 
for the case $n=2m = 2$ in the normal coordinate form. 
\be \label{phix21}
\phi_X (x_{2\p},x_{1\p},x_2,x_1) =
x_{1 2\p}^{2h_1-2h_2-1} x_{1\p 2}^{-1} x_{1\p 2\p}^{-2h_1} x_{12}^{2h_2} 
  \left (\frac{ x_{1\p 2\p} x_{12}}{x_{1\p 2}x_{1 2\p} } \right )^{\e} 
\ee

We define the integral operator as in (\ref{defQ}) 
 specifying $\mathbf{x}_I$ to $x_I$ and 
fix the integration
 in $x_{1\p} , x_{2\p}$ as  Pochhammer contours about the points $x_1, x_2$. 
 The functions $x_{12}^n $ are
eigenfunctions with the eigenvalues in terms of the Euler Beta function,
\be \label{BB}
 B(-\e, n+1-2h_1+ \e) B(1+n-2h_2, 2h_1-2h_2-\e). \ee
Comparing the dependence on $n$ with (\ref{REW})
we confirm that this operator with the kernel (\ref{phix21})
represents the $R$ operator obeying the $RLL$
relation (\ref{RLL}), where 
the  spectral parameter difference as the argument of $R$ as in (\ref{RLL})
(\ref{Qv})  is $w = u-v= \e -h_1+h_2 $. 
 
For real  values of the positions  $x_1, x_2$ and 
for generic values of the parameters
the contours can be contracted to
the interval  between $x_1, x_2$.
Then we can  write the kernel  in the
simplex integral form. 
\be \label{JX}
\int_0^1 \mathrm{d}^3\alpha\, \delta(\alpha_1+\alpha_2+\alpha_3 -1) 
\chi(\alpha )
\delta(x_{11\p} - \alpha_1 x_{12}) \delta (x_{22\p} + \alpha_2 x_{12}) 
 = J_{\chi}. \ee  
This form is the convenient one for comparison with the evolution kernels of
generalized parton distribution as given in \cite{BFL85}, \cite{DK01}. 
It is also convenient
because the Fourier transformation is easily done. 

We notice that the weights of the parton fields as they appear in the action
(\ref{S2Bj}) correspond by (\ref{2h2l}) to the negative helicity value for
both physical states. The substitution rule for the helicities in the
kernels has the peculiarity, that the choice of the helicity sign depends of
the form of the integration in the action of the operator. The symmetry
demands that the helicities of the points to be integrated over, 
e.g. $x_{1\p} $,  in both
factors are opposite, $h_{1\p}, -h_{1\p}$, but there are two options. 
If the choice is oriented
on the  weights in the action, then the negative value is to be
taken in $\psi$ irrespective of the physical helicity state.

We consider first the case $h_1=h_2=-h$  for $h=1, \half$ and
show that  the evolution kernels for gluon $\to$ gluon
and quark $\to$ quark without helicity flip are reproduced.

In the limit $\e\to 0$ this $R$ operator kernel (\ref{phix21}) 
is proportional to 
$\delta(x_{12\p}) \delta(x_{2 1\p}) $ representing the operator of 
point permutation $P_{12}$. The operator $P_{12} R_{12}(\e) $ has the kernel
obtained from the above (\ref{phix21})
by the  permutation of the points $x_1$ and $x_2$.

The next term in the expansion of the latter operator 
in $\e$ is the relevant contribution $J^{(2h)}_{11\p} + J^{(2h)}_{2 2\p}$,
represented as the simplex integral (\ref{JX})  with the integrand
$$ \chi_{1 1\p} (\alpha, h) + \chi_{2 2\p} (\alpha), \ \ \ 
\chi_{1 1\p} = (1-\alpha_2)^{2h} \delta(\alpha_1)  $$
The next relevant contribution is obtained at $\e=-1$ as
$J_{11,2h} $, represented as the simplex integral with the integrand
$$ \chi_{11h} (\alpha) = \alpha_3^{2h-1}. $$

In the gluon case the third relevant contribution is obtained at $\e=-2$,
$J_{221}$, represented as the simplex integral with the integrand
$$ \chi_{22 1} (\alpha) = \alpha_1 \alpha_2. $$.

Thus the evolution kernel is obtained in the gluon case
as 
$$ W_{1 -1 1 -1 }(x_{2\p}, x_{1\p}, x_2,x_1) =
a_0 (J^{(2)}_{11\p} +  J^{(2)}_{22\p}) + a_1 J_{112} + a_2J_{221}, $$
and in the quark case as

$$ W_{\half -\half \half -\half} (x_{2\p}, x_{1\p}, x_2,x_1) =
b_0 (J^{(1)}_{11\p} +  J^{(1)}_{22\p}) + b_1 J_{111}. $$
The coefficients are not fixed by the symmetry, we can get the information
from the comparison with the parton splitting  results as discussed in the
next section. The eigenvalues are calculated from (\ref{REW}), (\ref{BB})
with the corresponding substitutions and the resulting dependence on $n$
is compatible with the  one of the anomalous dimensions calculated 
by (\ref{anom}) from the
corresponding parton splitting probabilities $W_{1 1}$ and
$W_{\half, \half}$. 

In the case $h_1 = -h_2 = - h$ we may substitute the negative value for both  
and obtain  $(J^{(2h)}_{11\p} +  J^{(2h)}_{22\p}) $ as the first non-trivial
term in the expansion in $\e$ at $\e=0$,
as  expected for the kernels of the transversity parton
distributions $W_{1 1 1 1 },  W_{\half \half \half \half} $
from the results of the Feynman rule calculations
 \cite{BFL85}, \cite{DK01}. 



The transition kernels are not obtained as specifications of
the YSC (\ref{Phi4--++v}), for these  we   have to start with
the crossed YSC (\ref{Phi4c}). The anlogous steps that lead above to 
(\ref{phix21}) result in 
\be \label{phi||}
\phi_{||} (x_{2\p},x_{1\p},x_2,x_1) =
x_{12}^{-1} x_{1\p 2\p}^{-1}   x_{1 1\p}^{2h_1-2h_{2\p}}
 x_{1 2\p}^{2h_{2\p}}  x_{1\p 2}^{-2h_1} 
\left ( \frac{x_{1 2\p}  x_{1\p 2} }{  x_{1 1\p}  x_{2 2\p} } \right )^{\e}
,
\ee
$ h_1 = -h_2, h_{1\p} = - h_{2\p}$.
We specify the contours and the parameter $\e$ to reproduce the 
transition kernels. 

We fix $\e= -2h_{2\p} -1 $ and let the contour of $x_{2\p}$ be a circle
around $x_1$. Then the kernel reduces to
$$ \delta(x_{12\p}) x_{12}^{2h_{2\p}} x_{1\p 2\p}^{2h_1} 
x_{1\p 2}^{-1-2h_1-2h_{2\p} }. $$

The action on functions of $x_1, x_2$ is defined  
as the Pochhammer contour about $x_1, x_2$ resulting 
on the basic eigenfunctions $x_{1 2}^n$ in the eigenvalues
proportional to 
$$ B(1+n+2h_1, -2h_1-2h_{2\p} ). $$
By comparing the $n$ dependence with the one of the anomalous dimensions
calculated from $W_{AB} (z)$ we see that 
for $h_{1} = -1 , h_{2\p} = -\half $ and $h_{1} = -1 , h_{2\p}= -1 $
the kernels for the transition gluon - quark $W_{1 -1 -\half \half} $
and gluon - gluon $W_{1-1 -11}$ with spin flip
are obtained. 
For $ h_{2\p} = -1, h_1 = -\half $ the transition kernel 
quark - gluon non-flip $W_{\half - \half 1 -1 }$ 
is obtained.  
Other similar choices  of the contours and the parameter $\e$ lead to
transition kernels with the same or opposite helicities.

\subsection{4-point YSC and evolution kernels in momenta}

The helicity form of the 4-point YSC reads at $n=2m=2$
\begin{multline*}
\Phi^{--++ \lambda} = \int \mathrm{d}^4c \, \varphi (c) \, 
\delta(\bar\lambda_1-c_{13}\bar \lambda_3-c_{14}\bar\lambda_4) 
\delta(\bar \lambda_2-c_{23}\bar \lambda_3-c_{24} \bar\lambda_4) 
\delta(\lambda_3 +c_{13}\lambda_1 + c_{23}\lambda_2) \\
\times \delta(\lambda_4 +c_{14}\lambda_1 + c_{24}\lambda_2). 
\end{multline*} 
We have again four delta distributions 
but they are not removing the
four $c$ integrations. We see the dependence between the related 
linear equations by multiplying the first by $ \lambda_1$, the second by
$ \lambda_2$, the third by $\bar \lambda_3$, the fourth by $\bar \lambda_4$, and
adding them with the result
$ \lambda_1 \bar \lambda_1 + 
\lambda_2 \bar \lambda_2 + \lambda_3 \bar \lambda_3
+\lambda_4 \bar \lambda_4 = 0. $

We use the first three $\delta$-distributions to do the integrations over 
$c_{13}, c_{14},
c_{23}$. Their values  are fixed at
$$ c_{13}^{(0)} = \frac{\lambda_1 \bar \lambda_1 +\lambda_4 \bar \lambda_4 +
c_{24} \bar \lambda_4  \lambda_2 }{ \bar\lambda_3  \lambda_1}, \ \ \ 
c_{14}^{(0)} = \frac{- \lambda_4 - c_{24}  \lambda_2}{
\lambda_1}, $$  $$
c_{23}^{(0)} = \frac{\bar \lambda_2 - c_{24} \bar \lambda_4}{\bar
\lambda_3}, \ \ \ 
c_{13}^{(0)}c_{24}- c_{14}^{(0)}c_{23}^{(0)}
 = \frac{ \bar \lambda_2  \lambda_4 + c_{24} (\lambda_1 \bar \lambda_1
+\lambda_2 \bar \lambda_2)}{\bar \lambda_3  \lambda_1}. $$

We change to
$ c = c_{24}  \lambda_2 \bar \lambda_4 $
and introduce $ k_I = \lambda_I \bar \lambda_I$ in order to separate the
one-dimensional momentum dependence.
$$  c_{13}^*= \lambda_1 \bar \lambda_3  c_{13}^{(0)} = k_1 + k_4 + c, 
\ \ \ 
c_{14}^*= \lambda_1 \bar \lambda_4 c_{14}^{(0)} = k_1+k_2+k_3 - c, $$ $$
c_{23}^*=  \lambda_2 \bar \lambda_3 c_{23}^{(0)} = k_2 - c, \ \ \ \ 
\lambda_1 \lambda_2\bar \lambda_3\bar \lambda_4 
(c_{13}^{(0)}c_{24}- c_{14}^{(0)}c_{23}^{(0)}) =
k_2 k_4 + c(k_1+k_2) $$
The
result of the correlator in helicity form is
 \be \label{Phi1234l} \Phi_{1234}^{--++ \lambda} 
=  
\lambda_1^{-2h_1} \lambda_2^{-2 h_2} \bar \lambda_3^{-2h_1} \bar
\lambda_4^{-2h_2} \int \mathrm{d}c \, \varphi(c^*) 
\ \ \delta(k_1+k_2+k_3+k_4).
\ee
The kernel of the $R$ operator obeying the RLL relation (\ref{RLL}),
(\ref{YBnormal})
is obtained by substituting $\varphi$ from (\ref{Phi4--++v})
and relabeling the points $1234 \to 2\p 1\p 2 1$. 
We omit the momentum conservation delta and the phase factor.
\be \label{Phi4X}
  \phi_{X 2\p1\p21, 4}^{--++} =
(k_{1\p} k_1 )^{h_1} (k_{2\p} k_2 )^{h_2}
\int dc (k_2\p + k_1 +c)^{-1- \e 2 h_1-2h_2} c^{-1-\e} 
(k_{1\p} k_1 + c(k_2\p +k_1\p))^{\e-2h_1}.   
\ee
The question about choosing the contour of the $c$ integral
simplifies in the forward case $k_2\p +k_1\p = 0 $. 
Then a non trivial kernel is obtained with
choosing the
Pochhammer contour about the branch points related to the first two factors
in the integrand. We obtain the kernel

\be \label{wX} 
  w_X(k_1, k_{1\p}; h_1, h_2,\e) )= (k_1k_{1\p})^{1-h_1+h_2 + \e}  \ \ \ 
(k_1-k_{1\p})^{2h_1-2h_2 -1-2\e }. \ee

In the case $h_1=h_2 = h$ this reproduces with the substitution
$k_1 = z, k_{1\p}= 1 $
contributions to the 
kernels $W_{AB}(z) $ of the particular physical helicities 
$h_A = h_B = \pm 1 = h, h_A = h_B = \pm \half = h$ by choosing the particular 
values  $\e = 0, -1, -2$.
$$ W_{1 1}(z) = W_{-1 -1}(z) =
2 z w_X(z, 1; 1,1, 0 ) + 4  z w_X(z, 1; 1,1, -1 ) + z  w_X(z, 1; 1,1, -2 ) = $$ $$ 
2 z w_X(z, 1; -1,-1, 0 ) + 4 z w_X(z, 1; -1,-1, -1 ) + z w_X(z, 1; -1,-1, -2
), $$
$$  W_{\half \half}(z) = W_{-\half -\half}(z) =
2z w_X(z, 1; \half, \half , 0 ) + z  w_X(z, 1; \half , \half, -1 )  = $$ $$
2 z w_X(z, 1; -\half , -\half , 0 ) +  z w_X(z, 1; -\half , -\half, -1 ). 
$$
The extra factor $z$ is included because the convolution 
integral is usually written as $\frac{dz}{z}$.

Similar to the scattering amplitudes the admitted values of $\e$ 
can be understood from the condition to have as singularities in
the scattering channels only simple poles. But here we see no
argument for the coefficients of these contributions. 

In the case $h_1=-h_2 = h$ (\ref{wX})  reproduces contributions to the 
evolution kernels of the anti-parallel helicity configuration
(projection $P_T $ in \cite{BFL85} ),
$$ w(z; 1,-1, 2) = w(z; \half, -\half, 1) = \frac{z}{1-z}, $$
which determine the scale dependence of transversity structure
functions and which do not fit into the parton branching picture.

Similar to the scattering amplitudes not all parton splitting kernels
can be reproduced with the 4-point YSC (\ref{Phi4--++}).
It takes the crossed YSC (\ref{Phi4c}) to reproduce 
$W_{AB}$ with helicity flip or transition from gluon to quark.


\be \label{Phi4||}
  \phi_{|| 2\p1\p21,4}^{--++} =
(k_{1} k_2 )^{-h_1} (k_{1\p} k_{2\p} )^{h_{1\p}}
\int dc (k_2  +c)^{- \e+ 2 h_1-2h_{2\p}} c^{-1} (k_1+k_2 +c)^{-1} 
(k_{1\p} k_1 + c(k_2\p +k_1))^{\e+2h_{2\p}}.   
\ee
In the forward case, $k_1+k_2 =0$, parton splitting kernels are reproduced
e.g. by fixing $\e = -2h_{2\p} -1 $ and the contour as a circle around the 
pole of the last factor in (\ref{Phi4||}). We choose as independent variables
$ k_1, k_1\p, h_1, h_{1\p}$ and obtain
$$ w_{||}( k_1, k_{1\p}, h_1, h_{1\p}) =
(k_1-k_{1\p})^{2h_{1\p} - 2h_1 -1} k_{1\p}^{-2h_1\p} k_1^{2h_1}. $$
Then 
$ z w(1,z;h_1,h_{1\p})  $ reproduces with $h_1= -1, h_{1\p}= +1 $ the 
gluon-gluon helicity flip and with $ h_1= -1, h_{1\p}= +\half $ the
gluon - quark helicity flip kernel.
$  z w(z,1;h_1,h_{1\p})  $ with $h_1= + \half, h_{1\p} = +1$
reproduces the quark - gluon non-flip kernel. 

Summarizing, we have seen that  the evolution kernels of the generalized
parton distributions are obtained from 4-point YSC. 
In the momentum representation their explicit form appears involved and
simplifies in the forward limit. In this respect the representation
in the light-ray positions discussed in the previous subsection
is more convenient.


\section{Application to the Regge asymptotics}

\subsection{Three-point YSC and the triple reggeon vertex}

The BFKL equation summing the leading contributions to the exchange of two
reggeized gluons is formulated in the transverse momenta for which the
complex number notation is convenient. The kernel of this equation has been
shown to be composed from the triple reggeon vertices  (\ref{V-++}) with the
propagators $\sim |k|^{ -2} $ for the internal lines as discussed in subsect.
2.3.

Consider the collinear triple vertex (\ref{M3}) which has been
obtained  in sect. 5.1  by the three-point YSC in helicity form for $n=2m=2$. 
We observe that the reggeon vertices are
obtained from two factors $M_3$ (\ref{M3}), one with the 
momenta $k_I$ substituted by
the complex variable $\kappa_I$ and the other with
the  momenta $k_I$ substituted by
the complex conjugated variable $\kappa_I^*$. 
The helicities are to be substituted in accordance with the
weights of the reggeized gluon being $(\ell_I, \bar \ell_I) = (0,0)$.
This means in both factors $h_1=h_2=+\half, h_3 = - \half $.

\be \label{VM3M3}
 \mathcal{V}^{-++} (\kappa_1, \kappa_2) =  \mathcal{V}^{+--} (\kappa_1,
\kappa_2)=
M_{3}(\kappa_1,\kappa_2, -\kappa_1-\kappa_2; \half, \half, -\half)  
M_{3}(\kappa_1^*,\kappa_2^*, -\kappa_1^*-\kappa_2^*; \half, \half, -\half).
\ee

\subsection{The factorized $R$ operator and the BFKL hamiltonian}

We consider the $L$ operator in the case $n=2$ after projection to the weight
$2\ell$ in normal coordinates $x= \frac{x_1}{x_2} $ (\ref{Lv1v2}). 
The matrix $L(v^{(1)}, v^{(2)})$ factorizes as 
\be \label{Lratio} 
L(v^{(1)},v^{(2)}) =
\begin{pmatrix}
  v^{(2)}+1 + x \dd &  \dd  \\
x (- x \dd + 2\ell)  & v^{(1)} - x \dd 
\end{pmatrix} = \ \  
v^{(1)} \hat V^{-1}(v^{(1)}) \hat D \hat V( v^{(2)}),  \ee
$$ \hat V(v) = 
\begin{pmatrix}
v & 0 \\
-x & -1 
\end{pmatrix}, \ \ 
\hat D = 
\begin{pmatrix}
1 & - \dd \\
0 & 1 
\end{pmatrix}.
$$
We summarize the parameter  permutation method for constructing the
Yang-Baxter $R$ operator \cite{SD05, DKK07}.
Consider the monodromy of two such $L$ operators,
\be \label{L1vL2v}
 L_1(v_1^{(1)}, v_1^{(2)} ) L_2(v_2^{(1)}, v_2^{(2)} ) \ee
and look for operators representing the elementary permutations
of the parameter sequence $v_1^{(1)}, v_1^{(2)}, $ $v_2^{(1)}, v_2^{(2)}$
in action on the monodromy, e.g.
$$ \mathcal{S}_{12}(v_2^{(1)}- v_1^{(2)})  L_1(v_1^{(1)}, v_1^{(2)} ) L_2(v_2^{(1)}, v_2^{(2)} ) 
=  L_1(v_1^{(1)}, v_2^{(1)} ) L_2(v_1^{(2)}, v_2^{(2)} )  \mathcal{S}_{12}(v_2^{(1)}-
v_1^{(2)}). $$
Using the factorized form  (\ref{Lratio}) we find
$$ \mathcal{S}_{12}(v) = x_{12}^v. $$
For the analogous permutation of the first two ( $ I=1$) or the last two
($I=2$) entries in the sequence we have 
$ \mathcal{S}_{11}(v_1^{(1)}-  v_1^{(2)})$ and $ \mathcal{S}_{22}(v_2^{(1)}-
v_2^{(2)})$ with
$$  \mathcal{S}_{II}(v) = x_I^{-v} \frac{\Gamma(x_I \dd_I +1) }{ \Gamma(x_I \dd_I
+1-v)}. $$
Consider the permutation of the sequence $  v_1^{(1)}, v_1^{(2)}, v_2^{(1)}, v_2^{(2)}$
to $ v_2^{(1)}, v_2^{(2)},  v_1^{(1)}, v_1^{(2)}$. It appears in the $RLL$
relation with the complete Yang-Baxter operator (\ref{RLL}) modified by the
 the point permutation, $\check \R_{12} = P_{12} \R_{12} $.
To construct it in terms of the $\mathcal{S}_{IJ}$ we
 write a corresponding
product of elementary permutations and its representation 
in action on the monodromy (\ref{L1vL2v}). 
$$ \mathcal{S}_{12}(v_1^{(1)} - v_2^{(2)} )\mathcal{S}_{22}(v_1^{(2)} -
v_2^{(2)} )  \mathcal{S}_{12}(v_1^{(2)} - v_1^{(1)} ) \cdot 
\mathcal{S}_{12}(v_1^{(1)} - v_1^{(2)} )  \mathcal{S}_{11}(v_1^{(1)} - v_2^{(1)} )
 \mathcal{S}_{12}(v_1^{(2)} - v_2^{(1)} ) =
$$ $$
\frac{\Gamma(x_{12} \dd_2 +1+ v_1^{(2)}- v_1^{(1)}) }{ \Gamma( x_{12} \dd_2 +1+
v_2^{(2)}- v_1^{(1)} ) } 
\frac{\Gamma(x_{21} \dd_1 +1+ v_1^{(2)}- v_2^{(1)}) }{ \Gamma( x_{21} \dd_1 +1+
v_1^{(2)}- v_1^{(1)} ) }. $$
In terms of the spectral parameters and the weights we obtain 
a form of the complete $R$ operator for the case $g\ell(2)$ in terms of
the normal coordinate and derivative operators,
\be \label{Rn1}
\check \R(v_1-v_2; \ell_1, \ell_2)  = 
\frac{\Gamma(x_{21} \dd_2 - 2\ell_1) }{ \Gamma( x_{21} \dd_2  -\ell_1 -
\ell_2+ v_2- v_1 ) } 
\frac{\Gamma(x_{12} \dd_1 + v_1- v_2- \ell_1 - \ell_2) }{ \Gamma( x_{12} \dd_1
-2\ell_1  ) }
\ee
It is inconvenient,
a form with the operator arguments commuting with the generators $S_1^a+
S_2^a$ would be preferable. 

We notice that  on the highest weight states in
the tensor product of representations by polynomials in $x_1, x_2$
are represented as in (\ref{BB}) by powers of the differences $x_{12}^n$.
 They are   eigenvectors of the  operator (\ref{Rn1}). 
The convenient form with the same eigenvalues
is obtained as
\be \label{Rn2}
\R(v_1-v_2; \ell_1, \ell_2)  = 
\frac{\Gamma(- \hat m + v_1- v_2)  }{ \Gamma( - \hat m+  v_2- v_1 ) } 
\ee
in terms of the operator $\hat m$ related to the tensor Casimir 
(\ref{Casimir})
as $ \hat m (1-\hat m) = x_{12}^2 \dd_1 \dd_2. $

We recall that in the position representation the solution of the BKFL
equation is expanded  in the basis functions $ E^{n, \nu}$ (\ref{Ex}) 
factorising into 
holomorphic and anti-holomorphic factors 
with the representation labels $n$ taking all integer values and
$\nu$ taking all real values. The point $x_0$ labels the vectors 
of a particular representation. 

These holomorphic basis elements $E_m$ do not obey the highest weight
condition for finite $x_0$. The argument leading to the convenient form
(\ref{Rn2}) does not apply in the straightforward way.
However, for a fixed $x_0$ we can transform
the canonical pairs $x_1, \dd_1, x_2, \dd_2$ by a M\"obius transformation
of the coordinates,
\be \label{Mobius}
 x_1\p = \frac{1}{x_{10}}, x_2\p = \frac{1}{x_{20}}.  \ee
Then the generators ${S\p}_1^a, {S\p}_2^a $ constructed according to
(\ref{Sa})
in terms of the new canonical pairs (depending on $x_0$) are such that 
$E_m$ obey the highest weight condition,
$$  ({S\p}_1^- + {S\p}_2^-)  E_{m}(x_1,x_2;x_0) = 0, \ \ \  
({S\p}_1^0 + {S\p}_2^0)  E_{m}(x_1,x_2;x_0) = m   E_{m}(x_1,x_2;x_0). $$
Thus (\ref{Rn2} ) with $\ell_1= \ell_2= 0$  is a convenient form 
of the $R$ operator related to the BFKL pomeron. The BFKL hamiltonian
operator is obtained up to normalization 
as the first non-trivial term in its expansion in $\e = v_1-v_2$.
$$ H = \psi (\hat m) + \psi(1-\hat m  ) - 2\psi(1) $$
directly related to the eigenvalues (\ref{Omegag}) and the operators forms
$H_g$ (\ref{Hg1}), (\ref{Hg2}).  

This hamiltonian describes the
nearest-neighbour interaction in the multiple exchange of gluonic reggeons.
The decomposition of the R operator results in the set of commuting local 
 obserables of the corresponding closed  spin chain
\cite{TTF83}.

\subsection{4-point YSC and the dipole form of the BFKL kernel}

We consider the YSC at $N=4, K=2, n=2$ in the homgeneous coordinate form
as in sect. 5.3,  perform the  integrals over $c$ and
change to the normal coordinates 
separating the scale factor as in (\ref{phi}).  As discussed in
sect. 5.3  we obtain from (\ref{Phi4--++v}) the R operator kernel in the normal
coordinate form (\ref{phix21}). 

In sect. 3.6 we have considered integral operators
on functions of two n-dimensional position variables.  Here we consider
functions of two points $\mathbf{x}_1, \mathbf{x}_2$ 
in the complex plane and integral operators
$$ Q \psi(\mathbf{x}_1, \mathbf{x}_2 ) = \int d^2\mathbf{x}_1\p
\psi(\mathbf{x}_1\p, \mathbf{x}_2\p)
\  K(\mathbf{x}_2\p, \mathbf{x}_1\p, \mathbf{x}_2, \mathbf{x}_1).
$$
with the kernels composed of a holomorphic and an anti-holomorphic factor.
Both factors are derived from YSC of the form (\ref{phix21}).
In both factors we set
 $\ell_{1\p}=\ell_{2\p}=\ell$. 
 The position arguments are
substituted as  the complex $x_I, I=2\p,  1\p, 2,1$ 
in the first factor and as the
complex conjugated $x_I^*$ in the second. 
 $$ K(\mathbf{x}_{2\p}, \mathbf{x}_{1\p}, \mathbf{x}_2, \mathbf{x}_1) =
\phi_X(x_{1\p}, x_{2\p}, x_1,x_2)\phi_X( x^*_{1\p},  x^*_{2\p},  x^*_1,
 x^*_2) =
 \frac{|x_{1\p 2\p}|^{2(-1-2\ell-\e)} |x_{1 2}|^{2(1+2\ell-\e)} }{
|x_{1\p 2}|^{2(1-\e)} |x_{1 2\p}|^{2(1-\e)} }.  
$$
In the decomposition in $\e$ the leading $\e^{-1}$ term corresponds to the
kernel of the 
permutation operator $P_{12}$. The next  term in the $\e$ expansion is
$$|x_{1\p 2\p}|^{2(-1-2\ell)} |x_{1 2}|^{2(1+2\ell)} 
\left ( |x_{1\p 2}|^{2} \delta^{(2)} (x_{1 2\p})  +|x_{1 2\p}|^{2} 
\delta^{(2)} (x_{1\p 2}) - 
\delta^{(2)} (x_{1 2\p})\delta^{(2)} (x_{1\p 2}) 
\int d^2 x_3 |x_{1 3}|^{2} |x_{1 3}|^{2} \right ).
$$
 In the limiting case corresponding to the reggeized gluon exchange 
$\ell \to 0$ the action is defined on functions
vanishing at coinciding arguments $x_{1\p} = x_{2\p}$. The resulting kernel
 can be written as
$$  |x_{1 2}|^{2}\int \frac{d^2 x_3}{|x_{1 3}|^{2}|x_{23}|^{2} }
 \left ( \delta^{(2)} (x_{2 1\p}) \delta (x_{2\p 3})  
+ \delta^{(2)} (x_{1 2\p}) \delta (x_{1\p 3})
- \delta^{(2)} (x_{1 2\p}) \delta (x_{21\p} )
\right ) 
$$
This is the dipole (or M\"obius) form of the BFKL kernel (\ref{dipole}).

\section{Conclusions}

The L operator matrices with the Jordan-Schwinger presentation of
the $g\ell(n)$ algebra as its matrix elements can be regarded as the
basis of our approach. The monodromies built as ordered matrix products of
L operators enter the definition of the YSC.  The  properties of the
L operators result in relations for YSC and in methods of their
construction. 

By elementary canonical transformations different forms of the
L operators are obtained leading to different representation forms of the
YSC. The original JS form and the helicity form appear in the applications. 

The R operators obeying the Yang-Baxter RLL relation
involving products of two such L operators play a central role.
A particular solution for such a R operator has  the
form of a closed contour integral over a shift operator. It serves as the
tool of constructing YSC. Integral operators built by
particular 4-point YSC turn out to be solutions of the RLL relation.
In the case $n=2$ the spectrum of operators obeying  the RLL relation
and a further operator form of R have been studied and applied.  

The L operators projected to the action on homogeneous functions depend on
two parameters, the order N monodromy composed  of them and the
corresponding N-point YSC depend on the set of 2N parameters.  We choose these
parameters in a form, where the action of the R operators implies permutations 
of these parameters.  
  
In the applications the monodromy parameters are related to the particle
types and helicities. The YSC with generic parameter values turn to
physically relevant objects by specification of the parameters.

We have shown that the 3-point amplitudes, the building blocks of
the iterative BCFW construction of tree amplitudes, are specifications of
3-point YSC. We have explained that this iterative construction is in fact
the  convolution method of YSC construction. We have shown how
specifications of the parameters of 4-point YSC result in the $2 \to 2$ 
tree amplitudes of gluons and quarks. 

We have seen that the parton splitting amplitudes, appearing in the
collinear limit from the 3-point amplitudes, are obtained from specifications
of 3-point YSC. The 4-point YSC define as kernels integral operators obeying
relations of the RLL type. The latter allow to derive the eigenvalues of
those operators. We have shown how by the  specification of parameters in
4-point YSC and
integration contours the kernels of the Bjorken evolution equation
are obtained. This has been confirmed by showing that the corresponding 
eigenvalues and the anomalous dimensions are proportional.

The reggeon triple vertex decomposes in a holomorphic and an anti-holomorphic
factor, each of which is obtained as a specification of 3-point YSC.
The R operator obeying the RLL relation with the L operators specified to
the normal coordinate form of the $g\ell(2)$ case can be obtained 
by the parameter permutation method and contains the (holomorphic part of)
BFKL hamiltonian as the first non-trivial term in its decomposition. 
The integral operator form of this R operator is obtained by 
specifying a 4-point YSC as its kernel. For the action on functions of
two points in the transverse plane the corresponding R operator 
with two factors of such 4-point YSC as its kernel is constructed.
The  dipole form of the BFKL hamiltonian appears in the
decomposition of the latter R operator.

\end{document}